\documentclass[twocolumn]{aastex7}
\usepackage{amsmath}
\usepackage{graphicx}
\usepackage{hyperref}
\usepackage{nameref}
\usepackage{graphicx}
\usepackage{placeins}

\shorttitle{}
\shortauthors{}

\begin{document}

\title{Direct Abundance Maps and Radial Metallicity Gradients of two Galaxies at z$\sim$4-5 in the GARDEN Survey}

\shorttitle{Spatially-resolved direct abundances in GARDEN galaxies }
\shortauthors{Stanghellini et al.}

\author[0000-0003-4047-0309]{Letizia Stanghellini}
\affiliation{NSF’s National Optical-Infrared Astronomy Research Laboratory, 950 N. Cherry Ave., Tucson, AZ 85719, USA}
\email[show]{letizia.stanghellini@noirlab.edu}

\author[0000-0002-3838-8093]{Susan A. Kassin}
\affiliation{Space Telescope Science Institute, 3700 San Martin Drive, Baltimore, MD 21218, USA}
\affiliation{The William H. Miller III Department of Physics \& Astronomy, Johns Hopkins University, Baltimore, MD 21218, USA}
\email[]{kassin@stsci.edu}

\author[0000-0003-4196-0617]{Camilla Pacifici}
\affiliation{Space Telescope Science Institute, 3700 San Martin Drive, Baltimore, MD 21218, USA}
\email[]{camilla.pacifici@gmail.com}

\author[0000-0002-9288-9235]{Jane E. Morrison}
\affiliation{Steward Observatory, University of Arizona,
933 North Cherry Avenue, Tucson, AZ 85751, USA}
\email[]{janem@arizona.edu}

\author[0000-0001-5414-5131]{Mark Dickinson}
\affiliation{NSF’s National Optical-Infrared Astronomy Research Laboratory, 950 N. Cherry Ave., Tucson, AZ 85719, USA}
\email[]{mark.dickinson@noirlab.edu}

\author[orcid=0000-0002-1106-4881]{Ezra Sukay}
\affiliation{The William H. Miller III Department of Physics \& Astronomy, Johns Hopkins University, Baltimore, MD 21218, USA}
\email[]{esukay1@jhu.edu}  

\author[0000-0003-2434-2657]{Celia R. Mulcahey}
\affiliation{The William H. Miller III Department of Physics \& Astronomy, Johns Hopkins University, Baltimore, MD 21218, USA}
\email[]{cmulcah4@jhu.edu}

\author[0009-0001-4421-2377]{Louis E. Bergeron}
\affiliation{Space Telescope Science Institute, 3700 San Martin Drive, Baltimore, MD 21218, USA}
\email[]{bergeron@stsci.edu}

\author[0000-0001-9367-0705]{Michael W. Regan}
\affiliation{Space Telescope Science Institute, 3700 San Martin Drive, Baltimore, MD 21218, USA}
\email[]{mregan@stsci.edu}

\author[0000-0001-9262-9997]{Christopher N. A. Willmer}
\affiliation{Steward Observatory, University of Arizona, 933 North Cherry Avenue, Tucson, AZ 85751, USA} 
\email[]{cnaw@as.arizona.edu}

\author[0000-0001-6065-7483]{Benjamin J. Weiner}
\affiliation{MMT\/Steward Observatory, University of Arizona, 933 N. Cherry St., Tucson, AZ 85721, USA} 
\email[]{bjweiner@arizona.edu}

\author[0000-0002-5686-9632]{Nadia Dencheva}
\affiliation{Space Telescope Science Institute, 3700 San Martin Drive, Baltimore, MD 21218, USA}
\email[]{dencheva@stsci.edu}

\author[0000-0002-9402-186X]{David Law}
\affiliation{Space Telescope Science Institute, 3700 San Martin Drive, Baltimore, MD 21218, USA}
\email[]{dlaw@stsci.edu}

\author[0000-0002-6219-5558]{Alexander de la Vega}
\affiliation{University of California, 900 University Ave, Riverside, CA 92521, USA}
\email[]{alexander.delavega@ucr.edu}

\author[0000-0002-6610-2048]{Anton M. Koekemoer}
\affiliation{Space Telescope Science Institute, 3700 San Martin Drive, Baltimore, MD 21218, USA}
\email{koekemoer@stsci.edu}

\author[0000-0003-1949-7638]{Christopher Conselice}
\affiliation{Department of Physics and Astronomy, The University of Manchester, Alan Turing Building, Oxford Road, Manchester, M13 9PL, UK}
\email[]{conselice@manchester.ac.uk}

\author[0000-0003-2098-9568]{Jonathan P. Gardner}
\affiliation{Sciences and Exploration Directorate, NASA Goddard Space Flight Center, 8800 Greenbelt Rd, Greenbelt, MD 20771, USA}
\email[]{jonathan.p.gardner@nasa.gov}

\author[0000-0003-2775-2002]{Yicheng Guo}
\affiliation{Physics and Astronomy, University of Missouri, Columbia, MO 65211, USA}
\email[]{guoyic@missouri.edu}

\author[0000-0002-2165-5044]{Francois Hammer}
\affiliation{Observatoire de Paris, 61 Avenue de l'Observatoire, 75014 Paris, FR}
\email[]{hammer@obspm.fr}

\author[0000-0002-6586-4446]{Alina Henry}
\affiliation{Space Telescope Science Institute, 3700 San Martin Drive, Baltimore, MD 21218, USA}
\email[]{ahenry@stsci.edu}

\author[0000-0002-4884-6756]{Benne W. Holwerda}
\affiliation{University of Louisville, Department of Physics and Astronomy, 102 Natural Science Building, Louisville KY 40292, USA}
\email[]{bwholw01@louisville.edu}

\author[0000-0001-9187-3605]{Jeyhan Kartaltepe}
\affiliation{Rochester Institute of Technology, 1 Lomb Memorial Drive, Rochester, NY 14623-5603, USA}
\email[]{kartaltepe@rit.edu}

\author[0000-0003-1581-7825]{Ray A. Lucas}
\affiliation{Space Telescope Science Institute, 3700 San Martin Drive, Baltimore, MD 21218, USA}
\email[]{lucas@stsci.edu}

\author[0000-0002-2202-5415]{Mathieu Puech}
\affiliation{LUX, Observatoire de Paris, Université PSL, Sorbonne Université, CNRS, 92190 Meudon, FR}
\email[]{Mathieu.Puech@observatoiredeparis.psl.eu}

\author[0000-0002-9946-4731]{Marc Rafelski}
\affiliation{Space Telescope Science Institute, 3700 San Martin Drive, Baltimore, MD 21218, USA; Department of Physics and Astronomy, Johns Hopkins University, 3400 North Charles Street, Baltimore, MD 21218, USA}
\email[]{mrafelski@stsci.edu}

\author[0000-0003-4702-7561]{Irene Shivaei}
\affiliation{University of California, 900 University Ave, Riverside, CA 92521, USA}
\email[]{ishivaei@cab.inta-csic.es}

\author[0000-0001-5576-0144]{Charlotte Welker}
\affiliation{Department of Physics, New York City College of Technology, City University of New York, New York NY, USA}
\email[]{CWelker@CityTech.Cuny.Edu}

\author[0000-0000-0000-0001]{Xinfeng Xu}
\affiliation{Department of Physics and Astronomy, Northwestern University,
2145 Sheridan Road, Evanston, IL, 60208, USA; Center for Interdisciplinary Exploration and Research in
Astrophysics (CIERA), 1800 Sherman Avenue,
Evanston, IL, 60201, USA}
\email[]{xinfeng.xu@northwestern.edu}  

\author[0000-0003-3466-035X]{{L. Y. Aaron} {Yung}}
\affiliation{Space Telescope Science Institute, 3700 San Martin Drive, Baltimore, MD 21218, USA}
\email{yung@stsci.edu}

\begin{abstract}
We investigate galaxies in the GARDEN (Galaxies at All Redshifts Deciphered and Explained with the NIRSpec MSA) survey that exhibit auroral emission lines, enabling spatially resolved measurements of electron temperature and direct oxygen abundances. Two galaxies in this survey have spectra suitable for this analysis: CANDELS 8005 at z=3.794 and CANDELS 7986 at z=4.702. For both galaxies, we measure auroral and key nebular emission-line fluxes across their full extent, allowing direct-method oxygen abundance determinations in individual spatial pixels (spaxels). These observations demonstrate the viability of deep JWST/NIRSpec MSA spectroscopy for spatially resolved chemical analyses at high redshift, aided by weak nebular continua and low interstellar extinction. We derive global direct abundances of 12 + log(O/H) = 8.008$^{+0.025}_{-0.027}$ for CANDELS~8005 and 7.89$^{+0.027}_{-0.028}$ for CANDELS~7986. Emission-line diagnostics indicate neither galaxy hosts an active galactic nucleus. A first-order kinematic analysis suggests a potential merger in CANDELS~8005. The direct abundances are consistent with strong-line estimates based on our data and recent high-redshift calibrations. We build emission line, radial velocity, strong-line abundance indices, electron temperature, and direct abundance maps for both galaxies, thanks to the excellent spatial resolution. From the direct abundance maps we measure linear radial metallicity gradients of $-$0.111$^{+0.026}_{-0.025}$ dex kpc$^{-1}$ for CANDELS~8005 (statistically significant), and 
$-$0.0928$\pm$0.0880 dex kpc$^{-1}$ for CANDELS~7986, where the large uncertainties limit the significance of 
the result. These results provide a rare direct measurement of a radial metallicity gradient at $z>0$ from direct-method abundances, offering key observational support for inside-out galaxy growth with feedback-regulated chemical enrichment.
\end{abstract}

\section{Introduction}
Metal abundances are fundamental parameters in galaxies, essential for characterizing their age and evolution. They change with the time evolution of stars, are related to other fundamental parameters such as galaxy mass, and serve as the basis for determining galactic archaeology in galaxies both near and far \citep{2019igfe.book.....C}. The metallicity in galaxies can be assessed through gas-phase metal abundances or stellar abundances. The gas-phase metallicity in galaxies has been studied in the local Universe through the abundance of oxygen, the most abundant metal in galaxies. In H~{\sc ii} regions, oxygen is typically singly or doubly ionized, thus abundances of the O$^+$ and O$^{2+}$ ions are sufficient for metallicity assessment. The strongest emission lines for typical H~{\sc ii} physical conditions fall in the optical spectrum and are easily observable in low-redshift galaxies.

The expected strengths of ionic oxygen lines are highly sensitive to the plasma's physical conditions, which govern ionic emissivity. Consequently, the {\it direct method} \citep[e.g.,][]{2003ApJ...591..801K} is considered the most accurate approach for determining the gas-phase abundances in galaxies. This method involves plasma diagnostics to measure electron densities and temperatures ($N_{\rm e}$, $T_{\rm e}$).
Electron temperatures are measured from the auroral to nebular line ratio, with both lines being collisionally excited \citep[e.g.][]{1984ASSL..112.....A,2019A&ARv..27....3M}. Several plasma diagnostic lines, both for $T_{\rm e}$ and $N_{\rm e}$, also fall into the optical range in local Universe targets, thus their abundance analysis can be completed using a single spectral range, which should be deep enough to reveal the faint auroral lines. The auroral and nebular lines, such as those emitted by the N$^{+}$, O$^{2+}$, and other ions, have been successfully used for electron temperatures \citep[e.g.][]{1986ApJ...308..322K,2015ApJ...806...16B}. 

Until recently, abundances from the direct method were available only for nearby (z$\leq$0.1) galaxies, and have been essential to calibrate strong-line methods used everywhere else \citep[see][]{1979MNRAS.189...95P,2001A&A...369..594P,2008ApJ...681.1183K,2016MNRAS.457.3678P}. However, spatially-resolved direct-abundance measurements are rare because the auroral lines used for plasma diagnostics are often too faint for current-generation telescopes, even in nearby galaxies.  This type of study has been elusive at high redshifts due to the faintness of the auroral lines. 

The 2D direct abundance studies to date are limited to local galaxies. The availability of JWST spectroscopy has opened many possibilities for exploring auroral lines in distant galaxies, where one can use the known physics and analysis to determine abundances from emission lines observed in the 0.6-5 $\mu m$ range. Very soon after the initial observations, JWST spectroscopy yielded for the first time the auroral abundances in galaxies at $z \sim 1-2$ \citep{2022ApJ...936L..14P,2023ApJ...943...75S, 2023ApJ...958L..11S,  2023MNRAS.525.2087B, 2024ApJ...962...24S,2024ApJ...966..228R}, and also at much higher redshifts \citep[e.g.][]{2023MNRAS.518..425C,
2023MNRAS.518..592K}.

The currently available direct abundances of galaxies with $z>0$ are based on spectra integrated over the entire galaxies. However, local Universe studies have shown that the spatial distribution of abundances in star-forming galaxies gives much more information about their formation and evolution than the simple 1D analysis. Abundance analyses of H~{\sc ii} regions in local star-forming galaxies \citep{1992MNRAS.259..121V,2010ApJ...723.1255R,2020ApJ...893...96B} reveal that the galactic metallicity is not uniform, in particular, the radial metallicity gradient from the galactic center to the periphery of the disks is almost invariably negative (higher abundances near the galactic center) and shallow. \citet{2014A&A...563A..49S} and \citet{2015MNRAS.450.3254P} examined large numbers of local Universe galaxies to determine that their radial oxygen gradients are remarkably similar and shallow. 

Spectroscopy of gravitationally lensed z$\sim$2 galaxies initially offered a different view, with highly positive gradient slopes \citep{2010ApJ...725L.176J}. As the sample sizes of observed z$\sim$2 lensed galaxies increased, gas-phase radial metallicity gradients were observed to be, for the most part, negative and shallow \citep{2000MNRAS.312..813S,2016ApJ...820...84L, 2013ApJ...765...48J}. Additional work at z$>$3 led to the discovery of even flatter gradients \citep{2014A&A...563A..58T}, which is consistent with steepening of radial metallicity gradients with time. Extending the direct analysis to H~{\sc ii} regions at $z>0.1$ would allow us to directly compare the models of galactic chemical evolution to data at different epochs of galaxy formation. 

With the advent of cosmology-based chemical evolutionary models of galaxies, the evolution of metallicity gradients became an even more important constraint to be determined from observations. The gradient evolution in cosmological times encodes the combined effects of stellar evolution, gas inflow/outflow, and accretion processes, which determine whether the gradient flattens or steepens with time since galaxy formation. Even sophisticated cosmological, chemical evolution models of spiral galaxies cannot give such constraints {\it ab initio} \citep{2011MNRAS.415.1469R,
2012A&A...540A..56P,2013A&A...554A..47G} and observational constraints are necessary to advance the field.

The evolution of radial oxygen gradients can be constrained with observations of stellar populations of different ages in the same galaxy, or by comparing radial oxygen gradients in galaxies at different redshifts. Since direct radial gradients are available to date only for local galaxies, previous studies have used strong-line indices for distant galaxies. Abundance indices are based on correlations between the flux ratios of strong emission lines and the oxygen abundances, and such correlations are derived by calibration of direct abundance data points. The strong-line abundance indices may have uncertain calibrations, or could be degenerate, with two possible abundance solutions for the same index value (e.g., R3), resulting in large uncertainties in the derived abundances. If the absolute uncertainties are larger than the radial gradient slopes, they become only marginally useful for constraining the models, and it has become a priority to seek spatial resolution in auroral abundances in high-redshift galaxies. Recent advances have improved these calibrations considerably, expanding their realm of application \citep[e.g.][]{2024A&A...681A..70L}.

Spatially-resolved, direct abundance analysis is feasible for a few galaxies in GARDEN (Galaxies at All Redshifts Deciphered and Explained with the NIRSpec MSA, GO-2123 Cycle 1, Kassin et al., in preparation), due to the unique depth of these observations. GARDEN galaxies are observed in {\it slitlet-stepping mode} to obtain spatially-resolved spectroscopy in 2D. In this paper, we show the early results from the analysis of two galaxies, CANDELS~8005 and CANDELS~7986 (CANDELS: The Cosmic Assembly Near-infrared Deep Extragalactic Legacy Survey, \citet{2011ApJS..197...35G,2011ApJS..197...36K}; CANDELS GOODS-S catalog, \citet{2013ApJS..207...24G}). Basic data and measured parameters for these two galaxies are given in Table~\ref{basic_data}. Strong line analysis of the GARDEN galaxies will be presented in a separate paper, including the discussion of index validity at these redshift ranges.

Throughout this paper, we assume a flat $\Lambda$CDM cosmology with H$_0$ =70 km s$^{-1}$ Mpc$^{-1}$ and $\Omega_M$ =0.3, and $\Omega_\Lambda$ = 0.7 \citep[e.g.][]{2011ApJS..192...18K}.

\begin{table*}[ht]
\caption{Physical properties of the analyzed galaxies}

\begin{tabular}{llll}
\hline
{Parameter}& {CANDELS~8005} & {CANDELS~7986}  \\
\hline
RA [deg]~\tablenotemark{a} & 
53.13888206  &   53.13647396\\
DEC [deg]~\tablenotemark{a}& -27.83544487& -27.83576167\\
$R_{\rm eff}$ [$\arcsec]$~\tablenotemark{b}&   0.7702& 0.7767\\

log$(M_{\star})$ [M$_{\odot}$]~\tablenotemark{c}&  8.913 $\pm$ 0.1 & 9.953 $\pm$ 0.1\\
log(SFR) [M$_{\odot}$/yr]~\tablenotemark{c}& 2.352 $\pm$ 0.3& 1.204 $\pm$ 0.3\\

log(SFR) [M$_{\odot}$/yr]~\tablenotemark{d}&  1.898 $\pm$ 0.031 &1.787 $\pm$ 0.043\\  

\hline
&&\\
1D analysis~\tablenotemark{e}\\
\hline
redshift~\tablenotemark{f}&  3.794$ \pm$ 1.08$\times10^{-5}$ & 4.702 $\pm$ 2.73$\times10^{-5}$      \\

$\sigma_{\rm V}$ [km s$^{-1}$]&    77.205 $\pm$ 0.842 &42.915 $\pm$ 0.334\\

$T_{\rm e}$ [K]&            13861 $\pm$ 291    &   15273 $\pm$ 382    \\

$N_{\rm e}$ [cm$^{-3}$] &     144 $\pm$ 32    & 285 $\pm$ 41    \\

log(O/H)+12, direct&       8.008$^{+0.025}_{-0.027}$ & 7.89$^{+0.027}_{-0.028}$ \\

log(O/H)+12, strong-line~\tablenotemark{g}& 8.173$^{+\dots}_{-0.422}$ & 7.899$^{+0.643}_{-0.357}$\\

log(O/H)+12, strong-line~\tablenotemark{h}&  8.006 $\pm$ 0.1 &7.881 $\pm$ 0.1\\
\hline
&&\\
2D analysis~\tablenotemark{e}\\
\hline
$\Delta$($T_{\rm e}$/10$^3$)/$\Delta R_{\rm G}$ [K/kpc]& 1.547$^{+0.48}_{-0.49}$&  1.369$^{+1.283}_{-1.368}$\\ 

log(O/H)+12, range&   7.68 - 8.19& 7.26 - 8.32\\ 
$\Delta$log(O/H)/$\Delta R_{\rm G}$[dex/kpc]& -0.111$^{+0.026}_{-0.025}$\tablenotemark{i}& -0.0928 $\pm$ 0.088  \\ 
log(O/H)+12, $R_{\rm G}$=0&  8.089$^{+0.027}_{-0.027}$& 7.964$^{+0.073}_{-0.075}$\\ 

\hline
\end{tabular}

\tablenotetext{a}{Gaia DR3 coordinates.}
\tablenotetext{b}{Radial sizes from \citet{2012ApJS..203...24V}.}
\tablenotetext{c}{From CANDELS catalogs \citep{2016ApJ...832...79P, 2023ApJ...944..141P}.}
\tablenotetext{d}{This work, see text.}
\tablenotetext{e}{S/N$>$3 in all diagnostic lines.}
\tablenotetext{f}{From the H$\beta$ and [\ion{O}{3}]~$\lambda\lambda$4959,5007 line fitting.}
\tablenotetext{g}{R2 Index (see Tab. 3) from \citet{2024ApJ...972..113G}, z=5, 4 for CANDELS~7986 and 8005 respectively.}
\tablenotetext{h}{R2 Index from \citet{2024ApJ...962...24S}.}
\tablenotetext{i}{This is the gradient measured with the exclusion of the problematic spaxel, see text. The gradients for all spaxels with S/N$>$3 is -0.103$^{+0.026}_{-0.028}$.}
\label{basic_data}
\end{table*}

\section{GARDEN Data Inspection and Sample Selection}\label{sec:observations}

\subsection{Observations and data reduction}\label{sec:obs_redux}

GARDEN obtained spatially-resolved, multi-object spectroscopy of 56 high-redshift galaxies using JWST's NIRSpec \citep{2023PASP..135f8001G,2022A&A...661A..80J, 2023PASP..135c8001B} and its micro-shutter array \citep[MSA,][]{2022A&A...661A..81F}. The observations are fully described in Kassin et al.\ (in preparation), with relevant details summarized here. GARDEN galaxies are observed with grating G235H and filter F170LP with nominal spectral resolution R$\sim$3000, and employ a slitlet-stepping mode, in which a combination of telescope offsets and MSA reconfigurations is used to move the micro-shutter apertures across the extended structures of the target galaxies, to spectroscopically map their 2D emission. GARDEN slitlet-stepping covers different spatial extents in the dispersion direction for different target galaxies, depending on their angular sizes. Both galaxies analyzed in the present work have small sizes ($\lesssim$0.5\arcsec), and the MSA stepping for these two objects mapped a total extent of approximately 0.35\arcsec\ in the dispersion direction by 3.0\arcsec\ in the cross-dispersion direction. 

\begin{figure*}[ht]  
    \centering
    \includegraphics[width=\textwidth]{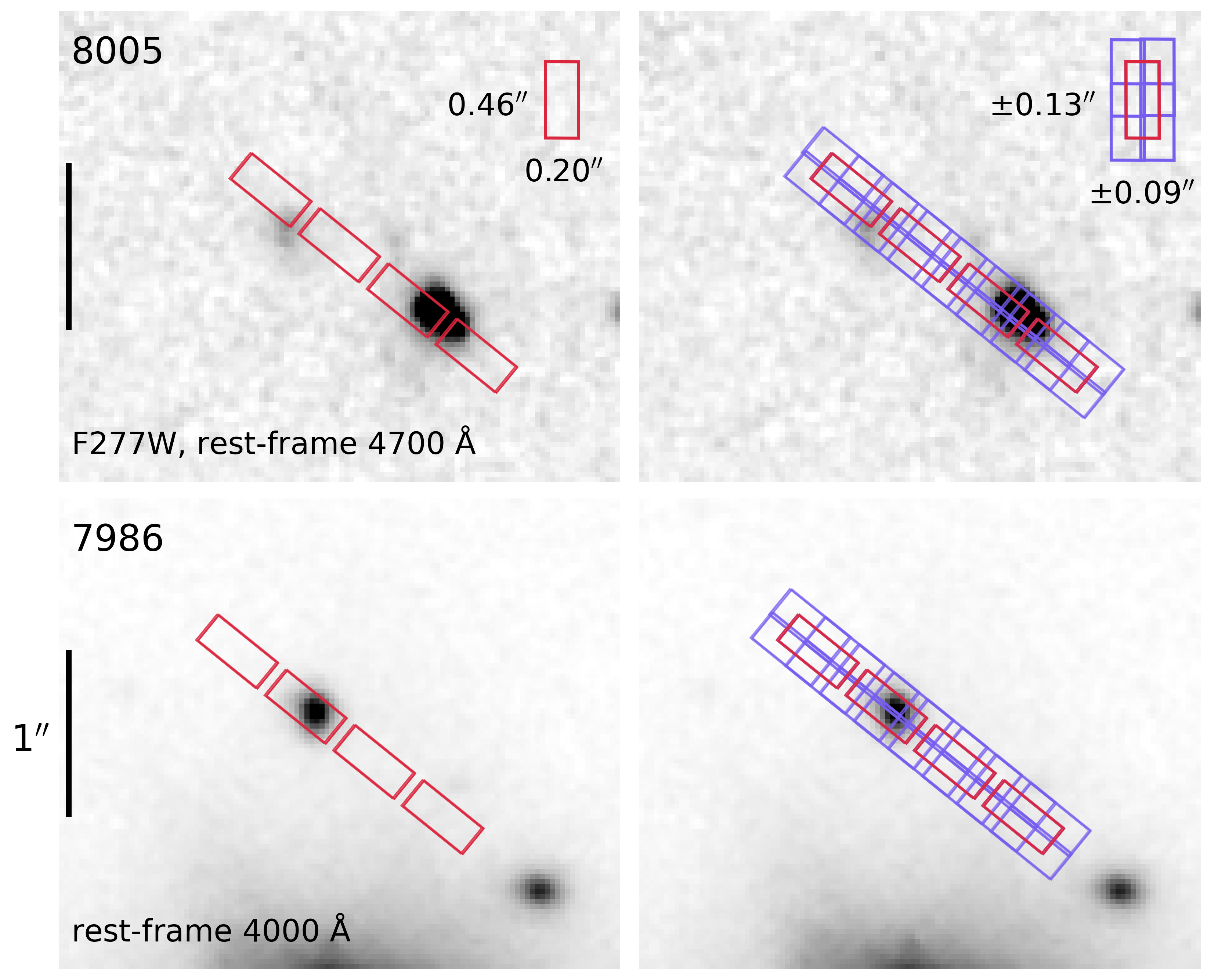}
    \caption{CANDELS~8005 (top row) and 7986 (bottom row) were observed by repeating a single slitlet step on the galaxy six times. There are five unique locations for the slitlet that make up the dither pattern. The central slitlet is overlaid (in red) on the F277W image in all figures. The slitlets in the four dithers (in purple) are shown on the right images. The dithers move out to the corners, offset in the dispersion direction by $\pm$0.09 arcseconds, and by $\pm$0.13 arcseconds in the cross-dispersion direction.}
    \label{slits}
\end{figure*}

The effective exposure time varies across the spectral cube as a result of the dither and stepping pattern. The majority of the data ($\sim 60\%$) has exposure times between 24 and 38 hours, while the remaining regions of the cube are shallower, with exposure times ranging from approximately 2 to 24 hours. In Figure \ref{slits}, we show the slit positions for the two galaxies analyzed in this paper. For each galaxy, the slitlet consists of four shutters, with one to two shutters placed over the galaxy and the remaining shutters positioned on the local sky. At each step, a five-point dither pattern is employed to sample across the vertical gaps between shutters and to mitigate the effects of bad pixels.

\begin{figure*}[ht]
\centering
\includegraphics[width=\textwidth]{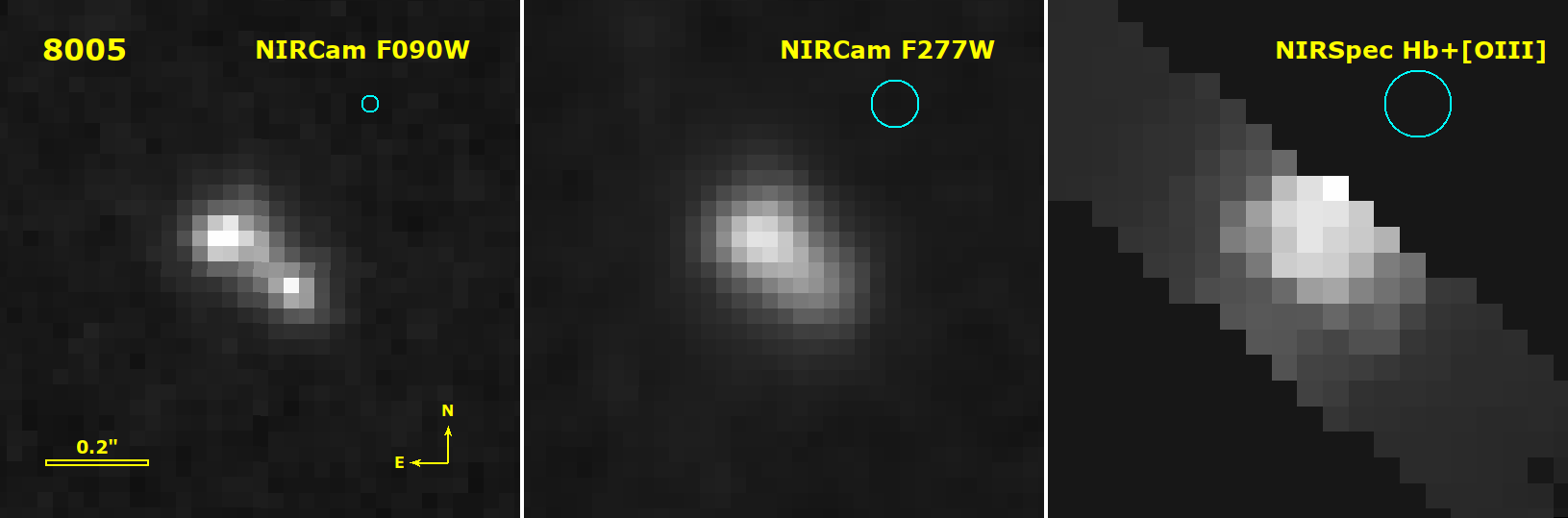}
\caption{Images of CANDELS-8005. Left and center:  NIRCam images in F090W and F277W, corresponding to rest-frame central wavelengths of 1900 and 4700 \AA~ respectively, for z=3.794.  F090W samples rest-frame ultraviolet emission, while F277W measures rest-frame optical light.  The [\ion{O}{3}]~$\lambda$ 5007\AA\ and H$\beta$ emission lines fall into the low-transmission wings of the bandpass. The cyan circles, with diameter of 0.33 and 0.092\arcsec respectively, represent the PSF for the NIRCam filters shown. Right:  GARDEN NIRSpec slitlet-stepping map, summing emission from the H$\beta$ and [\ion{O}{3}] 4959,5007 emission lines plus nearby continuum. The cyan circle, with diameter of 0.13\arcsec, represents the approximate NIRSpec IFU PSF size in the wavelength range of our spectra \citep[see][their Fig. 7]{2024NatAs...8.1443D}.}
\label{8005_images}
\end{figure*}

\begin{figure*}[ht]
\centering
\includegraphics[width=\textwidth]{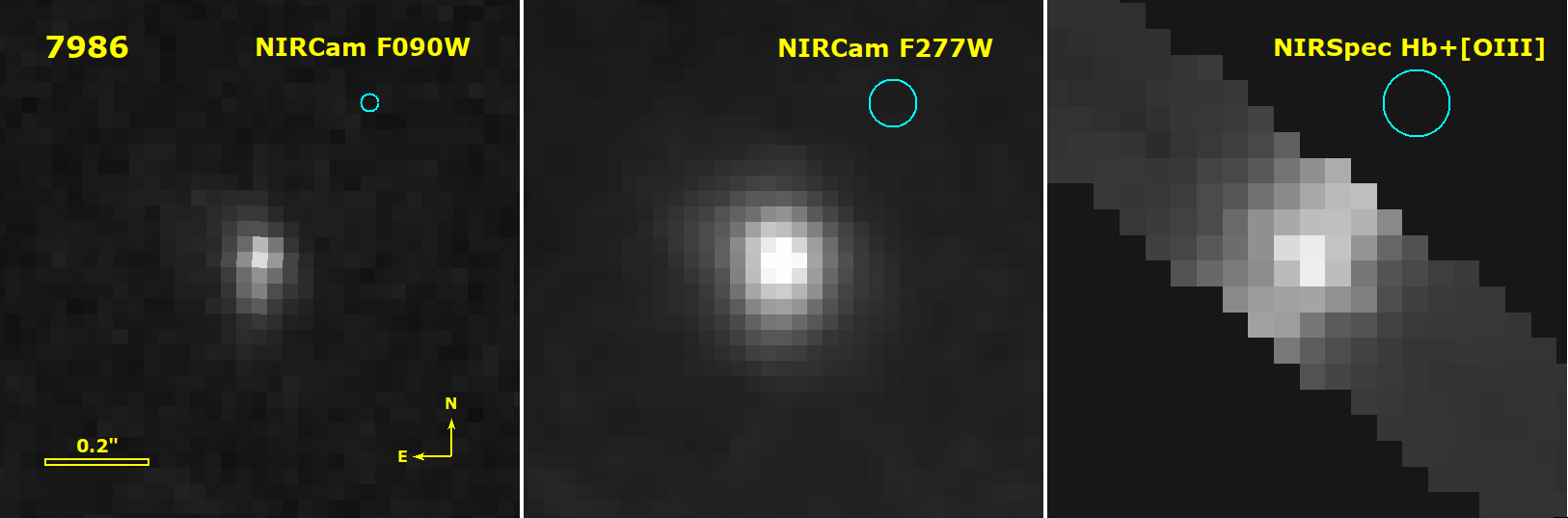}
\caption{Images of CANDELS~7986: Left and center:  NIRCam images of galaxy in F090W and F277W, or rest-frame central wavelengths of 1600 and 4000 \AA~ respectively, for z=4.702.  F090W samples rest-frame ultraviolet emission, while F277W measures rest-frame optical light, including the [\ion{O}{3}]~$\lambda$ 5007\AA\ and H$\beta$ emission lines.   Right:  GARDEN NIRSpec slitlet-stepping map, summing emission from the H$\beta$ and [\ion{O}{3}]~$\lambda\lambda$ 4959,5007 emission lines plus nearby continuum. The cyan circles represent the PSF for the three modes, with the same values as in Fig.\ref{8005_images}}.

\label{7986_images}
\end{figure*}

The spectra obtained at different spatial positions are reassembled to form a data cube of wavelength versus 2D position, analogous to the products of integral field spectroscopy.  We adopt the 3D drizzle algorithm described by \citet{2023AJ....166...45L} and adapted to the case of the NIRSpec MSA as described by Morrison et al. (in preparation).  In brief, this algorithm computes the overlap between the 2D detector pixels  and the 3D data cube voxels, composed of two spatial dimensions (spaxel) and one spectral dimension, and uses this fractional volume overlap together with a readnoise model to determine the flux for each data cube voxel from the weighted sum of the contributing detector pixels. The drizzle kernel uses square pixels of size 0.1\arcsec\ in both the dispersion and cross-dispersion directions, and an output spaxel size of 0.05\arcsec.

\subsection{Sample selection}\label{sec:sample_selection}

The GARDEN Survey Team has reduced the NIRSpec MSA data and produced data cubes for each of the 56 GARDEN galaxies (Kassin et al., in preparation). From these, we select a galaxy subsample where direct abundance analysis is feasible. We inspected all GARDEN galaxies data cubes initially with the {\it jdaviz} \citep{jdaviz_zenodo} routine {\it cubeviz} to look at the galaxy images in the light of the prominent emission lines.  
We then extract 1D, rest-frame spectra for the whole galaxies. We reassess the redshifts using the detection of strong unequivocal emission lines, such as the H$\beta$--[\ion{O}{3}] group. The measurements for direct oxygen abundance analysis include detecting the lines with a reasonable S/N ratio (e.g., above 3), and measuring their fluxes, of: (1) two or more Balmer lines, to measure the Balmer decrement and studying local extinction; (2) the strong metal lines emitted by the O$^{+}$ and O$^{2+}$ ions; (3) the auroral and corresponding nebular line pairs for electron temperature diagnostics; and (4) the diagnostic doublet for electron density determination. The ability to measure direct abundances is usually limited by the S/N ratio of the auroral lines, which are much fainter than other emission lines.

A general list of the diagnostic lines to search for in galaxies depends on the spectral range and the galaxy redshifts. Table~\ref{diagnostic} includes such diagnostic lines to search for in the GARDEN survey. In the Table, we list the excited ion, the line identification and rest-frame wavelength in the air (in \AA), the type of line emitted (i.e., whether it is a recombination line, RL, or a collisionally-excited line, CEL, and if it is an auroral or nebular CEL line), and the physical nebular parameter(s) that the line is diagnosing. The temperature diagnostics in the Table include the auroral lines for ions in ground configuration $p^2$. The [\ion{O}{2}] and [\ion{S}{2}] temperature diagnostics lines for ground configuration $p^3$ are not available in the GARDEN sample.

\begin{table*}[ht]
\caption{Diagnostic lines}
\label{diagnostic} 

\begin{tabular}{llll}
\hline
\colhead{Excited ion} & \colhead{Emission line, rest-frame $\lambda$}& \colhead{Description}\tablenotemark{a}& \colhead{Diagnostics}\\
\hline
H$^+$&     H$\gamma$ 4340&   RL& Balmer decrement, abundance (H$^+$) \\
H$^+$&      H$\beta$ 4861&      RL& Balmer decrement, abundance (H$^+$) \\
H$^+$&      H$\alpha$ 6563&   RL& Balmer decrement, abundance (H$^+$) \\
N$^+$&     [\ion{N}{2}] 5755& Auroral, CEL&    $T_{\rm e}$\\
N$^+$&  [\ion{N}{2}] 6548, 6584&   Nebular, CEL&     $T_{\rm e}$\\
O$^{2+}$&   [\ion{O}{3}] 4363& Auroral, CEL&    $T_{\rm e}$\\
O$^{2+}$&   [\ion{O}{3}] 4959, 5007&    Nebular, CEL&   $T_{\rm e}$, abundance (O$^{2+}$) \\
Ne$^{2+}$& [\ion{Ne}{3}] 3343& Auroral, CEL& $T_{\rm e}$\\
Ne$^{2+}$& [\ion{Ne}{3}] 3869, 3968& Nebular, CEL& $T_{\rm e}$\\
S$^{2+}$& [\ion{S}{3}] 6312& Auroral, CEL&   $T_{\rm e}$\\
S$^{2+}$& [\ion{S}{3}] 9069, 9531&  Nebular, CEL&  $T_{\rm e}$\\
O$^+$& [\ion{O}{2}] 3727, 3729&         Nebular, CELs   &     $N_{\rm e}$, abundance (O$^+$)\\
\hline
\end{tabular}

\tablenotetext{a}{RL: recombination line, CEL: collisionally-excited line.}
\end{table*}

We found that the direct oxygen abundance diagnostic is available in two of the 56 GARDEN galaxies, CANDELS~8005 and CANDELS~7986. For these galaxies, all key diagnostic lines—including the strong [\ion{O}{2}] and [\ion{O}{3}] lines, the Balmer lines, and the plasma diagnostic lines—are detected with S/N $>$ 3 in at least 2/3 of all the spaxels where H$\beta$ is measured, thus the abundance analysis is performed in $\sim$2/3 of the galaxy surface.  This enables spatially resolved plasma diagnostics and abundance analysis, including the characterization of radial gradients.

We note that the only available temperature diagnostic line in both galaxies is [\ion{O}{3}] $\lambda4363$. While this is sufficient for determining the O$^{2+}$ abundance, the O$^{+}$ emitting region may have a different temperature. This mismatch can introduce systematic uncertainties in the derived abundances \citep[e.g.,][]{2021MNRAS.501.3695C}, which we quantify below for each galaxy.

We show the NIRCam images of the galaxies studied in this paper, extracted from the JWST Advanced Extragalactic Survey (JADES) data release DR2 \citep{2023arXiv230602465E,2023arXiv231012340E}, in Figures \ref{8005_images} and \ref{7986_images}, where we also show their GARDEN data in a wavelength cut that includes H$\beta$ and the strong [\ion{O}{3}]~$\lambda\lambda$4959,5007 lines. All panels have the same spatial orientation (N up, E left).

The morphological differences between the NIRCam images, which trace the stellar continuum, and the NIRSpec images, dominated by gaseous line emission, arise from the convolution of the line emission with the MSA shutter pattern in the GARDEN Survey data cubes. This convolution imparts the characteristic boxy appearance to the NIRSpec reconstructions. The effect is particularly pronounced for CANDELS~7986, while it is less evident for CANDELS~8005, whose intrinsic morphology is already relatively boxy and comparable in size to a single MSA shutter.

\section{Data analysis} \label{sec:data_analysis}
\subsection{Spectral inspection, flux continuum, and line fitting}

\begin{figure*}[ht]
   \centering
    \includegraphics[width=\columnwidth]{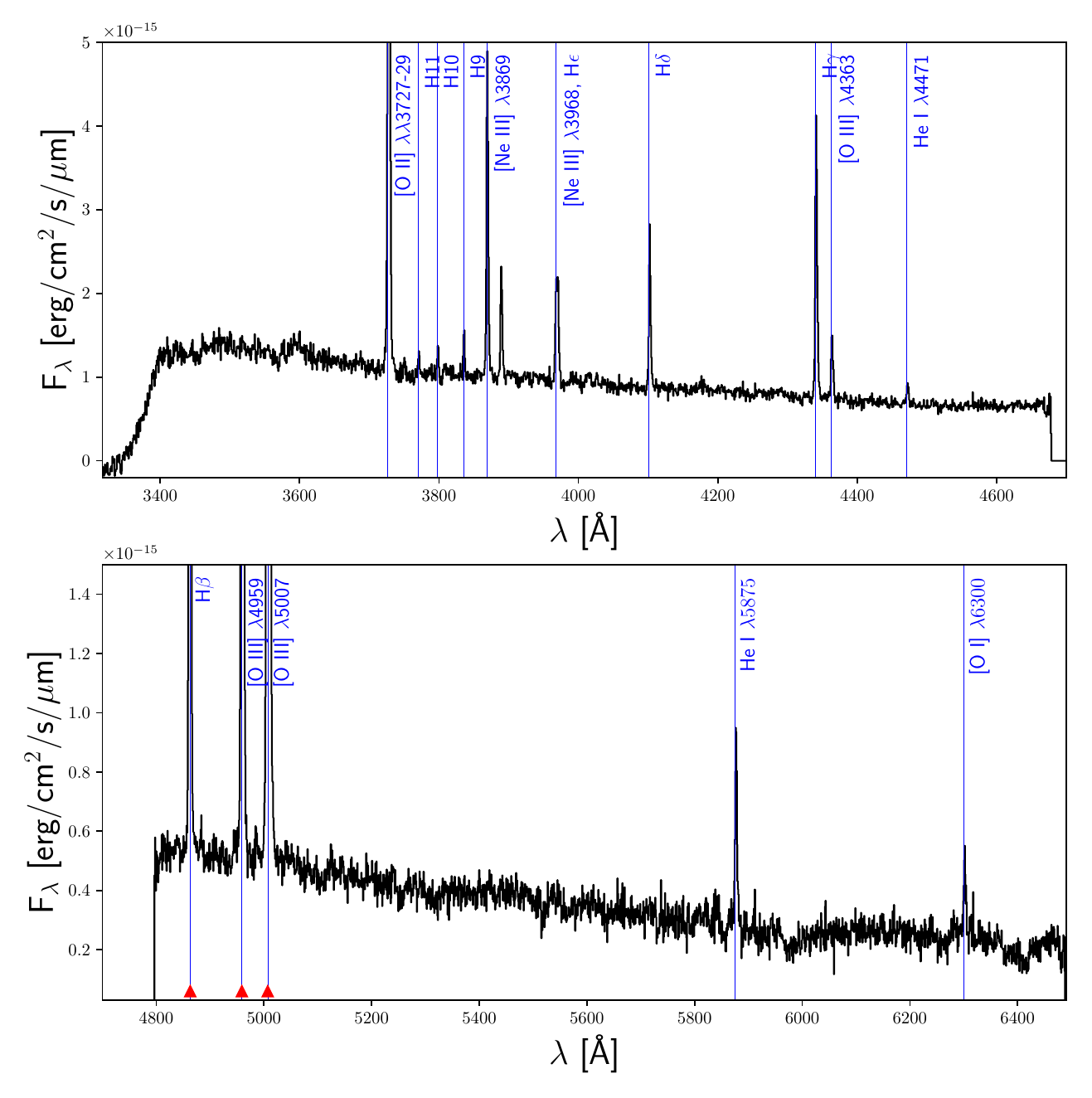}    
      \caption{1D spectrum of CANDELS~8005, flux density versus wavelength, obtained by spatially collapsing the cube over all spaxels inside the defined box (see text). Top and bottom panels are the spectral component on either side of the NIRSpec detector gap. The y-axis has been cut to show all lines, and in each panel the strongest lines above the y limit are indicated with red arrows. 
      The thin vertical lines and labels indicate the most prominent observed lines used for the analysis in this paper. 
}
              \label{8005_onedspec} 
   \end{figure*}
   
\begin{figure*}[ht]
\centering
    \includegraphics[width=\columnwidth]{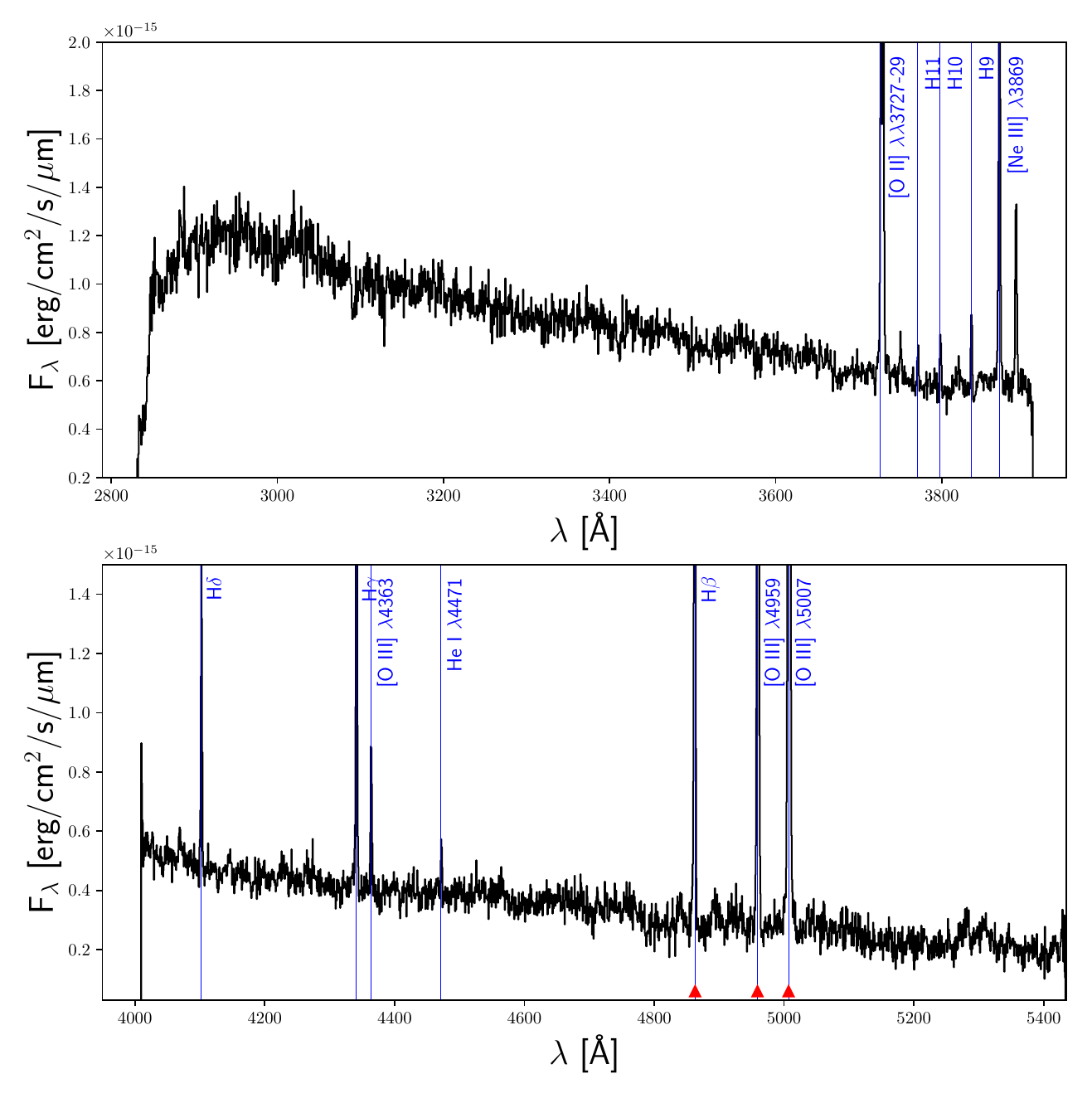}    
      \caption{Same as Fig.~\ref{8005_onedspec}, but for CANDELS~7986
}
              \label{7986_onedspec} 
\end{figure*}

Our analysis uses identical procedures for both galaxies. We study both the collapsed spectra for the whole galaxies, and the spectra of each individual spaxel. We use Gaussian fitting routines to measure the fluxes of all the relevant emission lines. We fit the emission lines individually or in groups (e.g., the [\ion{O}{2}] $\lambda\lambda$3727, 3729~\AA~ doublet, the H$\beta$ and [\ion{O}{3}] $\lambda\lambda$4959, 5007~\AA~lines, H$\gamma$ and [\ion{O}{3}] $\lambda$4363~\AA). The output of the fitting procedure includes the amplitude (A), the line dispersion ($\sigma_{\lambda}$), and the central wavelength for each emission line in each spaxel. 

The GARDEN survey data cubes give the surface brightness $SB_{\nu}$ in MJy/sr. The flux density per spaxel is $F_{\nu} {\rm [erg/cm^{2}/s/Hz/spaxel]} = 2.3504 \times 10^{-28} \times SB_{\nu} {\rm [MJy/sr]} \times \Omega {\rm [arcsec^2]}$, where $\Omega$ is the spaxel's solid angle (or area). We convert to $F_{\lambda} = c~F_{\nu}/\lambda^2$.  By expressing c in $\mu$m/s and $\lambda$ in $\mu$m we obtain:  

\begin{equation}
\begin{split}
F_{\lambda}  ~{\rm  [erg/cm^{2}/s/\mu m/spaxel] } =  \\
 (7.0464\times10^{-14})/\lambda^2 ~\times SB_{\nu} {\rm [MJy/sr]} \times\Omega {\rm [arcsec^2]}. 
\end{split}
\end{equation}

\begin{table*}[ht]
\caption{Flux measurements from the 1D spectra}
\label{indices} 
\begin{tabular}{rll}
\hline

\colhead{Emission line} &  \colhead{CANDELS~8005}\tablenotemark{a}& \colhead{CANDELS~7986}\tablenotemark{a}\\
\hline
H$\beta~\lambda$4861&   1.120$\times10^{-17}\pm6.338\times10^{-20}$& 5.166$\times10^{-18}\pm5.603\times10^{-20}$\\
$c_{\rm H\beta}$&             0.068 $\pm$ 0.028 &0.084 $\pm$ 0.040\\

\hline

[\ion{O}{2}]~$\lambda$3727&   51.489 $\pm$ 0.855 & 40.061 $\pm$ 0.673  \\

[\ion{O}{2}]~$\lambda$3729&    67.511 $\pm$ 1.469 & 47.659 $\pm$ 0.952 \\

[\ion{Ne}{3}]~$\lambda$3869&   29.253 $\pm$ 0.823&    $\dots$\\

[\ion{Ne}{3}]~$\lambda$3968&        46.209 $\pm$ 0.589&   50.891 $\pm$ 0.863\\

H$\delta~\lambda$4101&   23.91 $\pm$ 0.494   & 24.361 $\pm$ 0.739\\

H$\gamma~\lambda$4340 &      45.952 $\pm$ 0.518&  45.8 $\pm$ 0.666\\

[\ion{O}{3}]~$\lambda$4363&        10.642 $\pm$ 0.519&    13.665 $\pm$ 0.666\\

[\ion{He}{1}]~$\lambda$4471&        3.769 $\pm$ 0.301&    4.4882 $\pm$ 0.699\\

\ion{He}{2}~$\lambda$4686&          $\dots$&        2.164 $\pm$ 0.563\\

[\ion{O}{3}]~$\lambda$4959&        228.87 $\pm$ 0.566&    231.34 $\pm$ 1.084\\

[\ion{O}{3}]~$\lambda$5007&          680.3 $\pm$ 0.566&  700.45 $\pm$ 1.085\\

[\ion{He}{1}]~$\lambda$5875&         12.295 $\pm$ 0.450& 26.665 $\pm$ 0.977\\

[\ion{O}{1}]~$\lambda$6300&   5.858 $\pm$ 0.446&    12.708 $\pm$ 0.968\\

\hline
\end{tabular}

\tablenotetext{a}{H$\beta$ fluxes and errors are given in [erg cm$^{-2}$ s$^{-1}$], while all other fluxes and errors are in terms of $F_{\rm H\beta}$=100}.
\label{fluxes}
\end{table*}

The science derived from the data analysis is based on emission-line fluxes. There are two sources of uncertainties in the line fluxes: the native uncertainties of the observed data, which are recorded as data cube flux errors (see the GARDEN Survey paper, Kassin et al., in preparation), and the uncertainties in the Gaussian line fitting. We measured the contribution to the native data errors for all measured emission lines but with particular attention to the auroral [\ion{O}{3}] lines, our faintest diagnostic line. The auroral lines are fitted simultaneously with H$\gamma$, using the spectral range
$\lambda_{\rm obs}({\rm H}\gamma)-0.02 < \lambda < \lambda_{\rm obs}[{\rm O III}]+0.02 ~[\mu m]$. 
We measured both the error in the line that derives from the native flux errors, and the error derived from the difference between the lines plus continuum fit and the observed spectrum. We repeat these measurements for each spaxel of each observed galaxy, finding that the uncertainties derived from the native flux errors are smaller than the errors in the fits. The same type of error ratios are seen in all the other fitted lines, thus we focus on the Gaussian fit errors.

We use the scaling relations by \citet{1992PASP..104.1104L} to determine the S/N ratio:
\begin{equation} 
(S/N)_{\rm x} = C_{\rm x} ~\sqrt(FWHM/\Delta\lambda) (S/N)_{\rm 0}. 
\end{equation}

In this expression, the suffix ``x" stands for the parameters for which the scaling is performed, and the suffix ``0" refers to the maximum value of the parameter; FWHM=2.355$\times~\sigma$, where $\sigma$ is the standard deviation measured from the Gaussian fit, in $\mu m$, and $\Delta\lambda$ is the spectral dispersion in the wavelength axis (e.g., in $\mu$m/pixel). We can express $(S/N)_{\rm 0}=A/res$, where $A$ is the amplitude of the line, and $res$ is the median absolute deviation (MAD) of the residuals between the data and the Gaussian + continuum fit. We use C$_{\rm x}$=0.7 from \citet{1992PASP..104.1104L}, which is valid for both the integrated and maximum line flux cases, to obtain the signal-to-noise estimate in the flux:
\begin{equation}
(S/N)=0.7 (2.35 \sigma_{\lambda}/\Delta\lambda)^{0.5} (A/res), 
\end{equation}
which we use to calculate the local $S/N$ from the fit. 
\subsection{Reddening Correction, Plasma Diagnostics and Direct Abundance Analysis}

We analyze the plasma and metallicity of each galaxy both as a whole (1D analysis) and by spaxel (2D analysis), using the classical emission-line analysis. The analysis, based on nebular physics, is performed with the Python-based routines in {\it PyNeb} \citep{2015A&A...573A..42L,2020Morisset_Atom8} opportunely adapted to address our specific questions.

We begin with an assessment of the global galactic properties. The interstellar reddening due to the Galactic foreground is expected to be low in the direction of the GARDEN observations, which are in the Hubble Ultradeep Field \citep[HUDF,][]{2006AJ....132.1729B}. There may still be dust reddening within the individual galaxies, thus we measure the extinction by Balmer decrement analysis. Since we are working with emission lines, we use the extinction in the line approach. The logarithmic extinction constant in H$\beta$ is defined by the relation:
\begin{equation}
F_{\lambda} / F_{\rm H\beta} = I_{\lambda} / I_{\rm H\beta} 10^{- c_{\beta} (f_{\lambda}-f_{\rm H\beta})} 
\end{equation}
where $F_{\rm \lambda}$ and $F_{\rm \beta}$ are the observed line fluxes, $I_{\rm \lambda}$ and $I_{\rm \beta}$ are the intensities, or emitted fluxes, c$_{\beta}$ is the logarithmic extinction constant at H$\beta$, and $f_{\lambda}-f_{\rm H\beta}$ is the adopted extinction law \citep{1989ApJ...345..245C}. We apply this equation for H$\gamma$, and the first term of the equation is the observed Balmer ratio, while the intrinsic line ratios for the Balmer lines, $I_{\rm gamma}/I_{\rm beta}$, is from the line emissivity ratio calculated for the observed plasma parameters (case B, see PyNeb). 

Temperatures and densities are measured together and consistently, by minimized independent uncertainties in the line analysis. In this study we measure $T_{\rm e}$ and $N_{\rm e}$ initially from the observed emission lines, uncorrected for extinction, and then we reiterate and measure the plasma parameters once again with the corrected emission lines, if extinction is relevant.

The primary objective of this work is to measure elemental abundances and analyze their spatial variations within each galaxy using metallicity maps and gradients. Oxygen abundances are derived from the ionic abundances of O$^+$ and O$^{2+}$. The direct abundance calculation relies on atomic data models, which depend on local plasma conditions and the observed ionic emission lines. Consequently, the abundance error analysis incorporates uncertainties in the line fluxes of the strong oxygen lines,  and the uncertainties in plasma diagnostics, which are propagated from the uncertainties in the measurements of the diagnostic lines. To account for all sources of uncertainty at all stages of our analysis, we associate each emission line with a set of simulated observations that follow a Monte Carlo distribution within the observed flux uncertainties. We obtain the final results of the calculated physical parameters ($T_{\rm e}$, $N_{\rm e}$, and abundances) as the median of the Monte Carlo values, while the standard deviations give the uncertainties, including propagation. This methodology, detailed in the PyNeb manual, has been successfully applied in abundance analyses of planetary nebulae (PNe) and H~{\sc ii} regions \citep[e. g.][]{2015A&A...573A..42L,2024A&A...690A.264W}.

\section{Detailed analysis of CANDELS~8005}
\begin{figure*}[ht]
\centering
    \includegraphics[width=\textwidth]{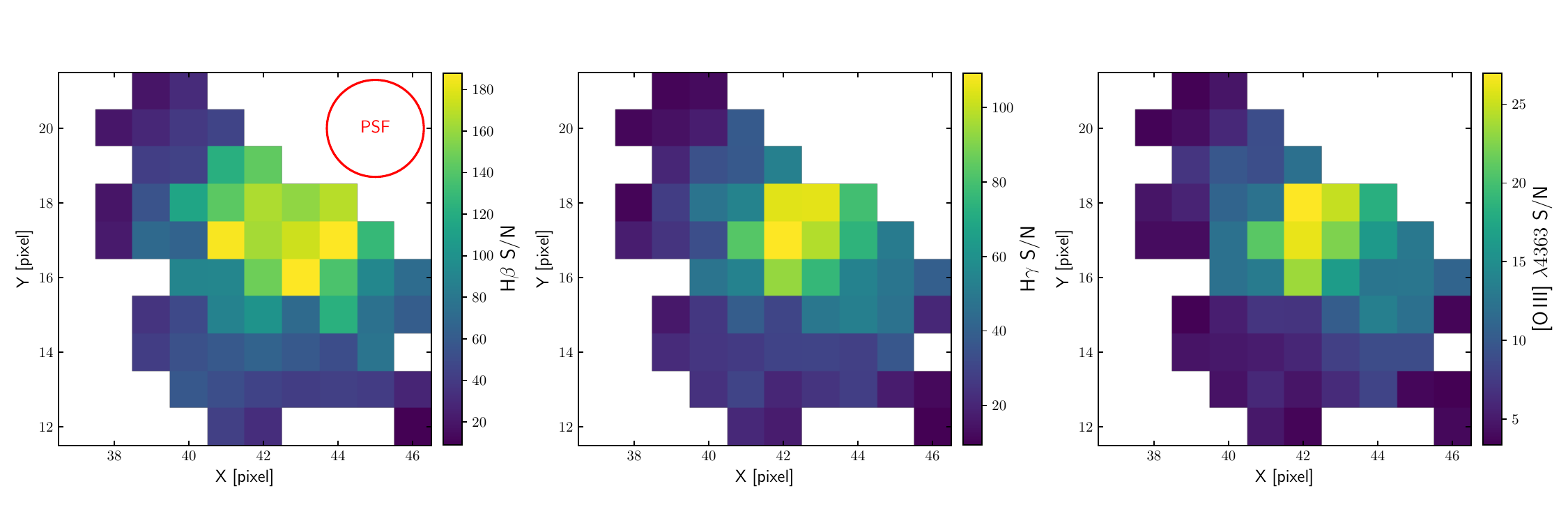}%
 \caption{CANDELS~8005: The S/N maps for the fluxes in H$\beta$ (left panel), H$\gamma$ (central panel), and [\ion{O}{3}] $\lambda$4363 (right panel), for the spaxels used in the following analysis (S/N$>$3 for all major diagnostic lines). In the left panel we indicate the approximate PSF size, 0.13\arcsec.
}
\label{8005_SN_map}
\end{figure*}

\subsection{1D spectral analysis}

In the collapsed 1D spectrum of CANDELS~8005 (Fig.~\ref{8005_onedspec}) we find 
the [\ion{O}{2}]~$\lambda\lambda$3727,3729 doublet, all Balmer series lines from H11 through H$\beta$, with H$\epsilon$ blended with [\ion{Ne}{3}]~$\lambda$3869, the nebular and auroral [\ion{O}{3}] lines, and the \ion{He}{1} emission lines at 
$\lambda$4471 and $\lambda$5875, and [\ion{O}{1}]~$\lambda$6300~\AA. Two high-excitation lines, \ion{He}{2}~$\lambda$4686 and \ion{Ne}{5}, are unfortunately non observable, the former would fall in the detector spectral gap, and the latter does not fall within the observed wavelength range of this galaxy. It is thus not possible to establish the galaxy ionization with precision. 

We measure the redshift of the galaxy as a whole by simultaneously fitting the H$\beta$ and strong [\ion{O}{3}] lines, obtaining $z=3.794$. From direct analysis of the collapsed spectrum we measure the major spectral lines. We calculate $T_{\rm e}$=12861$\pm$291 K, and $N_{\rm e}$=144$\pm$32 cm$^{-3}$. The nebular continuum predicted from these physical parameters, representative of typical H~II region conditions, is negligible and does not affect our emission-line analysis \citep{Osterbrock2006}. The extinction constant is measured with Eq. (4) from the observed F$_{\rm H\gamma}$/F$_{\rm H\beta}$ flux ratio, and the theoretical emissivity ratio of these lines computed with PyNeb using the measured temperature and density, adopting the extinction law of \citet{1989ApJ...345..245C}. We found $c_{\rm \beta}$=0.068$\pm$0.028, or E(B-V)=0.047$\pm$0.018, which is considered low extinction. 

It is worth noting that this extinction is likely to be dominated by dust internal to the galaxy rather than foreground extinction. The adopted extinction law is typically used for nebular attenuation. Other extinction laws \citep[e. g.][]{2000ApJ...533..682C} also give low extinction. We include the line extinction derived from the observed Balmer ratio in the analysis, however, including or excluding it in the line analysis produces negligible differences on the final science results. 

The total oxygen abundance for this galaxy is log(O/H)+12=8.008$^{+0.025}_{-0.027}$. The O$^+/$O$^{2+}$ abundance ratio of this galaxy is $\sim 15\%$, indicating that the impact of differing electron temperatures between the two ionization zones is limited. We derive star-formation rates (SFRs) from the extinction-corrected H$\beta$ luminosity. Observed H$\beta$ fluxes are corrected for dust attenuation. The H$\beta$ luminosities are converted to H$\alpha$ luminosities assuming Case~B recombination at an electron temperature consistent with the observed ones, for which $\rm H\alpha/\rm H\beta = 2.74$--2.75 \citep{Storey1995,Osterbrock2006}. Star formation rates are then computed using the calibration of \citet{Kennicutt2012}, appropriate for a Kroupa/Chabrier initial mass function. Uncertainties on the SFRs include propagation of measurement errors on the H$\beta$ flux and the extinction constant. but do not include any uncertainties carried by the calibration used. All measured parameters are in Table~\ref{fluxes}. 

\subsection{2D spectral analysis}

\begin{figure*}[ht]
\centering
\includegraphics[width=\columnwidth]{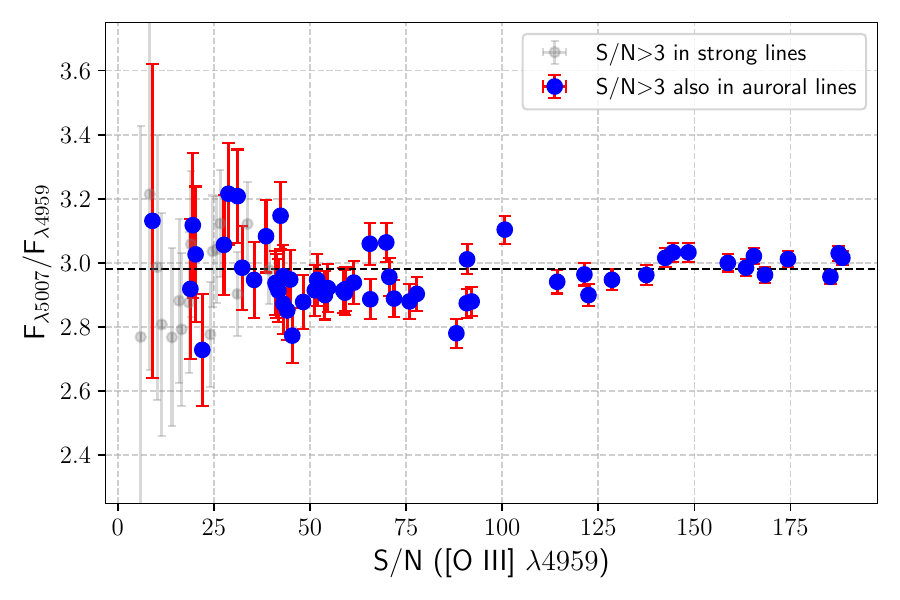}    
     \caption{CANDELS~8005: Scatter plot showing the observed [\ion{O}{3}] $\lambda$5007 to 
     $\lambda$4959 flux ratio, plotted against the signal-to-noise ratio of [\ion{O}{3}] 
     $\lambda$4959, for each spaxel. Blue symbols with red erorbars indicate the data for spaxels used in the abundance analysis. Grey symbols and error bars: Spaxels where the auroral lines were not detected above S/N$\sim$3, not used in the abundance analysis. The error bars are propagated from the Gaussian fit uncertainties. The clustering around 2.98, the theoretical ratio, demonstrates the consistency and reliability of the strong [\ion{O}{3}] line measurements across the spectral cube.}
\label{8005_O3ratio}
\end{figure*}
The projected image of the galaxy in our GARDEN data cube can be encompassed by a 10$\times$10 pixel box (0.5\arcsec$\times$ 0.5\arcsec), with 73 spaxels covering the galaxy H$\beta$ emission. We measured the line flux and uncertainty by Gaussian fits of the strong emission lines in each spaxel, and found S/N$>$3 in all the spaxels for the strong [\ion{O}{2}] and [\ion{O}{3}] lines, and Balmer lines. On the other hand, the auroral line [\ion{O}{3}]~$\lambda$4363 has S/N$>$3 in 57 of the 73 spaxels with signal. In Figure~\ref{8005_SN_map} we show the S/N ratio for all spaxels with S/N$>$3 in three main diagnostic lines, H$\beta$, H$\gamma$, and [\ion{O}{3}] $\lambda$4363~\AA.  The available data are adequate for our 2D analysis. 

To verify the quality and consistency of our emission line measurements, we examined the [\ion{O}{3}] $\lambda$5007 to $\lambda$4959 flux ratio for each spaxel in the spectral cube. This ratio is theoretically fixed at about 2.98 due to atomic physics \citep{2000MNRAS.312..813S}; significant deviations could indicate measurement or calibration errors.

Figure~\ref{8005_O3ratio} shows that even spaxels with relatively low S/N have ratios close to the theoretical value, in particular when we look at the spaxels where the electron temperature diagnostics are available. This suggests that the measurements of the strong [\ion{O}{3}] lines are robust throughout the spatial extent of the source, confirming that our spectral extraction and calibration methods reliably capture the spatially resolved emission line properties necessary for further physical diagnostics. On the other hand, the emission line errors from the fits are clearly underestimating the observed errors in our data set. We calculate the $\chi^2$ for the oxygen flux ratio, and its reduced value for the number of spaxels, $(\chi^2/(N-1))^{0.5}$=1.28 which is an approximate underestimation factor of our flux uncertainties. We will consider this result for our subsequent analysis.

\begin{figure*}[ht]
\centering
    \includegraphics[width=\columnwidth]{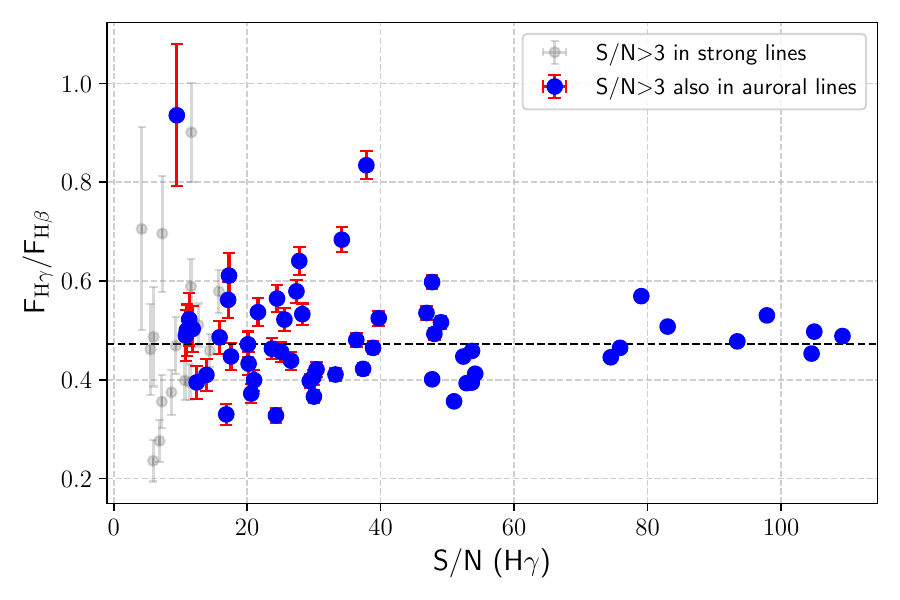}    
    \caption{CANDELS~8005: The observed Balmer ratio versus the S/N of the H$\gamma$ line in the galaxy spaxels. Spaxel symbols are as in Fig.~\ref{8005_O3ratio}. The horizontal broken line corresponds to the theoretical Balmer ratio, based on case B recombination (0.472) calculated for this galaxy.}
 \label{8005_ext}
\end{figure*}

\begin{figure*}[ht]
   \centering
    \includegraphics[width=\textwidth]{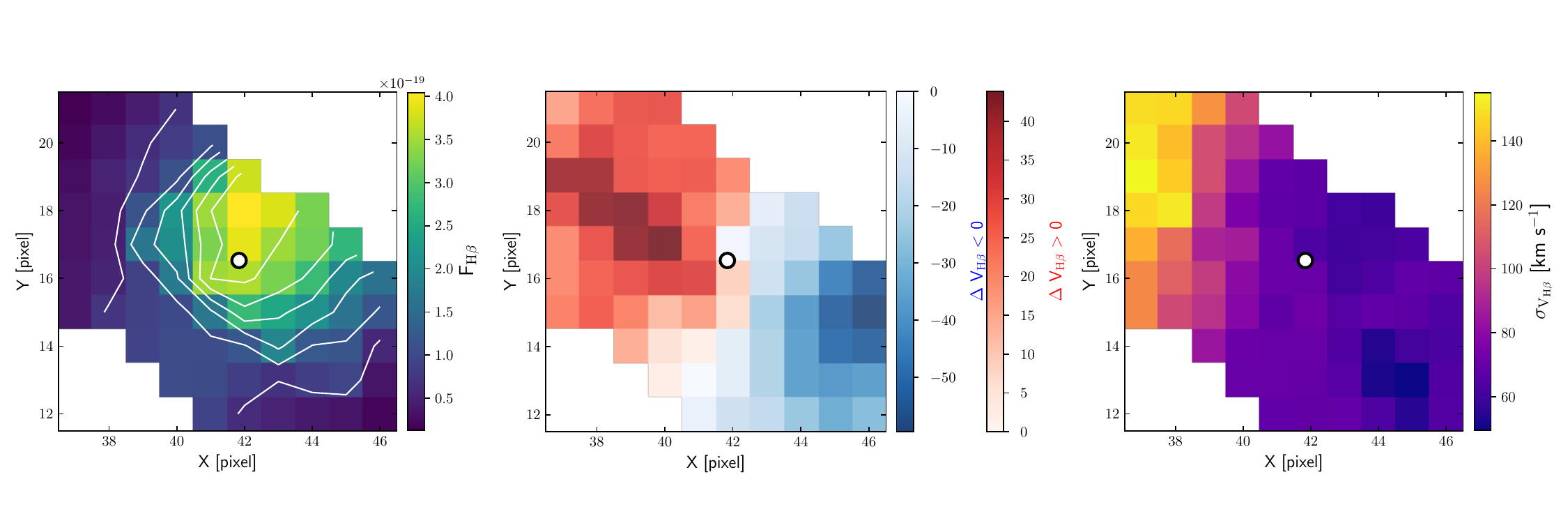}%
 \caption{CANDELS~8005: Spatial maps of selected emission line properties derived from spaxels with significant H$\beta$ detections (S/N $>$ 3). Left: H$\beta$ flux map, i.e., the Gaussian-fit of the H$\beta$ flux in the spaxels, with contour levels in white. The white dot is the flux centroid measured from the spectral extraction around H$\beta$, whose slightly offset location indicates galaxy asymmetry. Center: the velocity shift of the H$\beta$ line in each spaxel from that of the galaxy as a whole, where we plotted the positive and negative velocities with separate color maps, and the H$\beta$ centroid from the left panel analysis is shown for spatial reference. Right: velocity dispersion, in H$\beta$, corrected for the instrumental dispersion. Panels show only those spaxels with valid (S/N$>$3) measurements. The galaxy map is 10$\times$10 pixels (0.5$\times$0.5 \arcsec), with orientation as in Fig.~\ref{8005_images}.
}
         \label{8005_diagnostic_fbeta_dv_disp_map}
   \end{figure*}
The observed $F_{\rm H\gamma}/F_{\rm H\beta}$ ratio, shown in Figure~\ref{8005_ext}, reasonably clusters around the intrinsic case-B flux ratio calculated for this galaxy (i.e., for its observed plasma parameters), 0.472.  At low S/N the Balmer ratio has a lot of scatter, and there are high Balmer ratios in a few spaxels, suggesting that we likely observe data scatter rather than extinction. 

We know that the Balmer ratio decreases for the effect of dust absorption, or for hydrogen collisional excitation. The Balmer ratio may increase by starlight fluorescence \citep{2008ASPC..390..101L}. The lack of correlation between the Balmer ratio and $T_{\rm e}$ rules out collisional excitation as a significant factor. The data available do not allow further speculation, except that the extinction may be low in most spaxels. The 2 spaxels (coordinates (41,20), (46,1)) with auroral line detection (blue symbols) where the Balmer ratio is higher than 0.8 also have unrealistic oxygen line ratios, possibly due to cube-building issues (see $\S$5.2). We will single-out these spaxels in the gradient analysis (see $\S$4.3). 

\begin{figure*}[ht]
\centering
    \includegraphics[width=\textwidth]{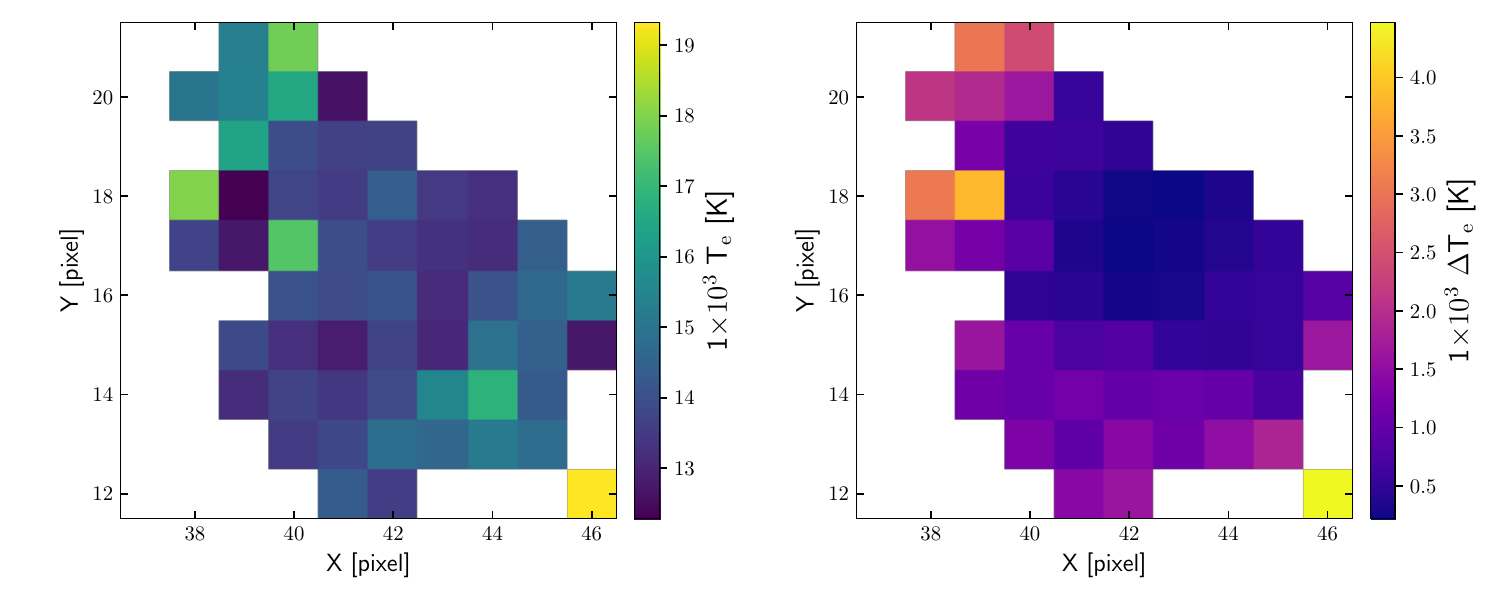}
 \caption{CANDELS~8005: Results of 2D mapping of spaxel temperature (left), and uncertainty, as derived from the Monte Carlo propagation of errors using the PyNeb plasma analysis (right). All spaxel values are derived from lines with S/N$>$3. The galaxy map is 0.5$\times$0.5 \arcsec, N is up, E is left.}
         \label{8005_diagnostic_Te_map}
\end{figure*}

\begin{figure*}[ht]
   \centering
    \includegraphics[width=\textwidth]{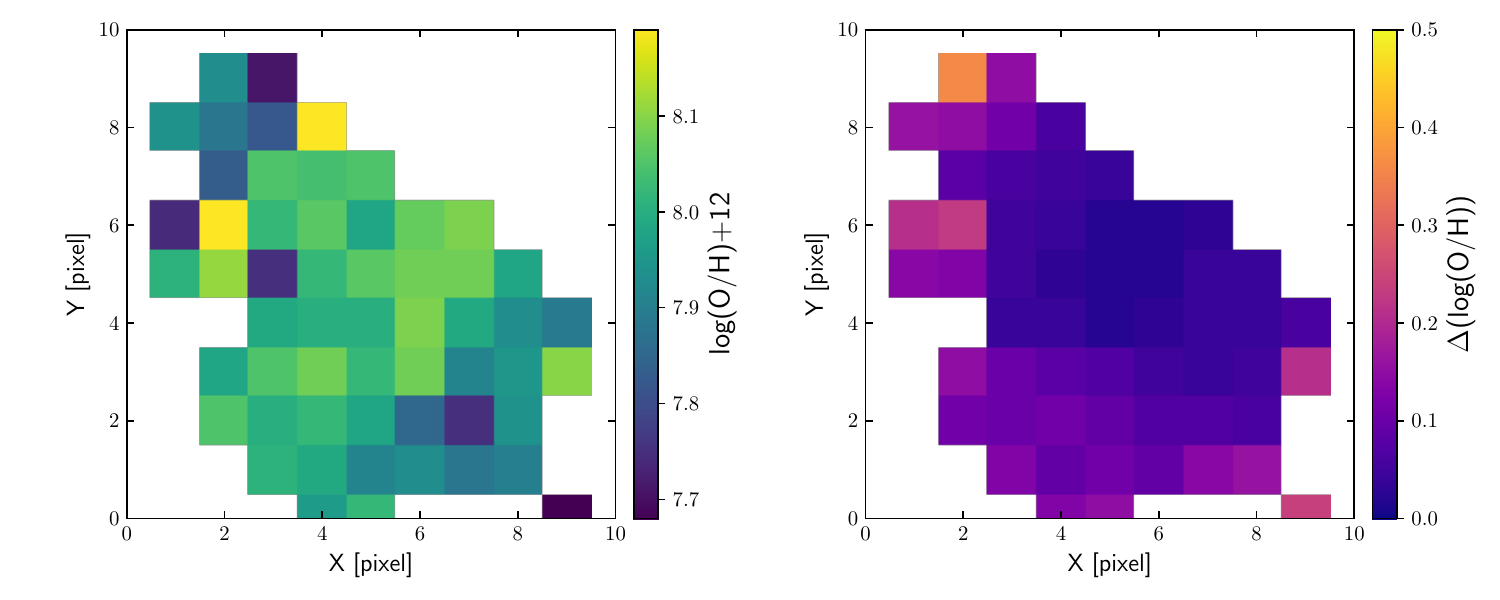}    
 \caption{CANDELS~8005: Results of 2D mapping of spaxel metallicity (in the usual format of log(O/H)+12, left) and uncertainty (right), as derived from the Monte Carlo propagation of errors using PyNeb. All spaxel values are derived from lines with S/N$>$3. The galaxy map is 0.5$\times$0.5 \arcsec, N is up, E is left.
}
\label{8005_diagnostic_OH_map}
\end{figure*}

In Figure \ref{8005_diagnostic_fbeta_dv_disp_map}, left panel, we show the H$\beta$ maps derived from line fitting in each spaxel. The central panel maps the velocity field $\Delta$V across the galaxy, and the right panel represent the calculated velocity dispersion corrected for the instrumental value of $47~\rm km~s^{-1}$. 
$\Delta$V is measured as the difference in radial velocity from the individual spaxel to the galactic systemic redshift, obtained using the collapsed 1D spectrum.  The radial velocity shows a structure suggestive of rotation of the galaxy around a NW-SE axis. It is worth noting that CANDELS~8005 deviates from a simple rotating disk because the gas kinematics are dominated by the velocity dispersion rather than rotation. Furthermore, the peak velocity dispersion is offset from the center, which is inconsistent with ordered rotating disks \citep[e.g.][]{2006ApJ...653.1027W}.
The structure of this galaxy as seen in the JADES images (Fig~\ref{8005_images}) at shorter wavelengths, clearly displays two separate blobs, which could be interpreted either as two separate star-forming regions within the same galaxy, or as two separate, possibly merging galaxies. The region where $\Delta$V approaches zero includes the zone of the brightest H$\beta$ spaxel. 

We limit the kinematic study in this paper to showing the velocity and dispersion maps for insight in the extinction structure of the galaxy. We found no correlations between the observed Balmer ratio and the radial velocity nor the dispersion, except that the data scatter of the Balmer ratio increases with $\Delta$V and $\sigma$V.

The electron density is measured from the [\ion{O}{2}]~$\lambda\lambda$3727, 3729 line ratio. Deblending the [\ion{O}{2}] lines is feasible in most spaxels. We verify by Monte Carlo simulations that using the density measured for the whole galaxy (Table~\ref{basic_data}) instead of the spaxel-specific one would not change the resulting abundances outside the given error bars. We found no evident correlations between the electron density and the velocity dispersion.

In Figure~\ref{8005_diagnostic_Te_map} we show maps of the electron temperatures and their uncertainties propagated from the combined line-fit errors. The maps show spaxels where S/N$>$3 in all emission lines involved in the analysis. We note a clear variation of the temperature across the galaxy. Since the oxygen abundance has strong dependence on the electron temperature, we expect to see spatial variation in the metallicity. In Figure~\ref{8005_diagnostic_OH_map} we show the metallicities, expressed as log(O/H)+12, and their uncertainties by spaxel, obtained by propagating from the abundance and the plasma analysis, for all spaxels where all diagnostic lines have  S/N$>$3. We measure low-uncertainty weak-line abundances in the central spaxels, while several spaxels in the periphery carry higher uncertainties, which we will propagate into the error analysis of the metallicity gradients.

We find no correlation between the extinction and the electron temperature and density, confirming that the observed variations of the Balmer ratio are probably due to error scatter.
   
\begin{figure*}[ht]
\centering
    \includegraphics[width=\columnwidth]{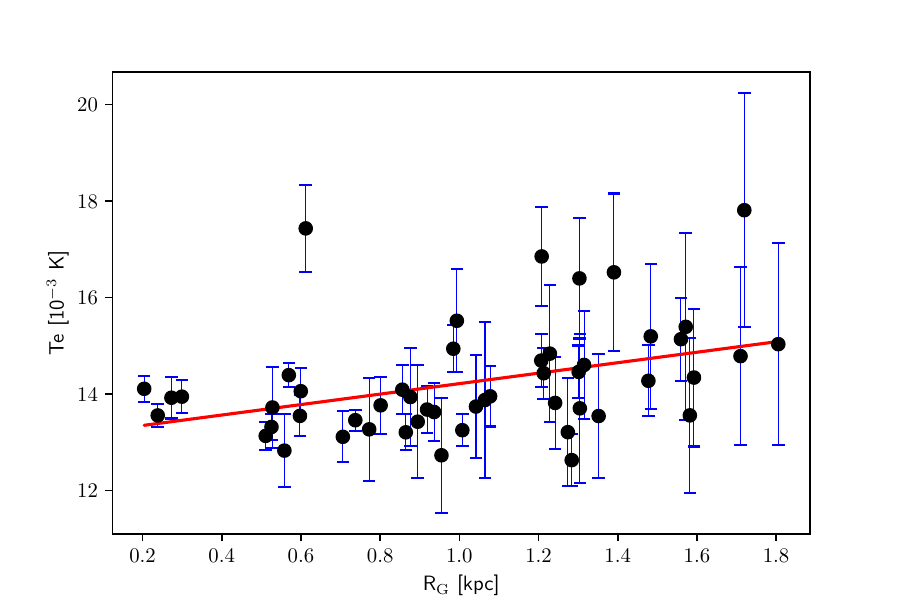}  
      \caption{CANDELS~8005: Electron temperature versus the linear distance from the galaxy center (i.e., the centroid of the H$\beta$ emission, see text), $R_{\rm G}$, plotted with the bootstrap linear fit, which gives the radial Te gradient given in Table~\ref{basic_data}. }
\label{8005_Tegrad}
\end{figure*}

\begin{figure*}[ht]
\centering
    \includegraphics[width=\columnwidth]{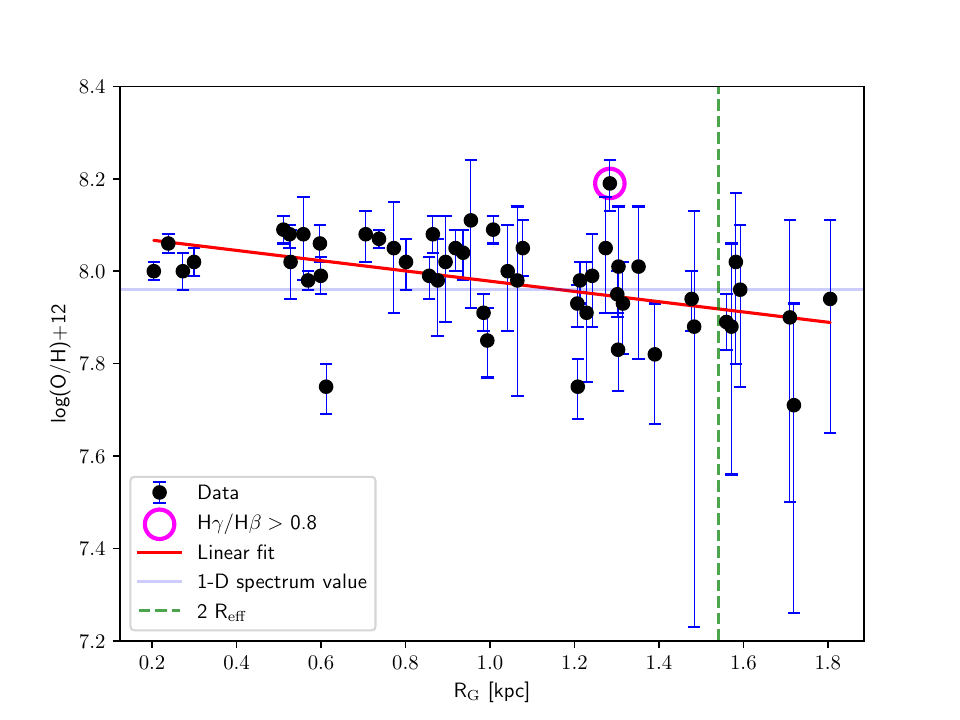}  
      \caption{CANDELS~8005: The metallicity in each spaxel, using the oxygen abundance as a proxy in the usual log(O/H)+12 format, versus the linear distance of the spaxel from the galaxy center (i.e., the centroid of the H$\beta$ emission, see text) $R_{\rm G}$. The black symbols are the measured values in each spaxel, and the red circle indicates the problematic data point, that has not been used in the fit (see $\S$4.2); the red line is the bootstrap linear fit, which gives the radial metallicity gradient (in Tab.~\ref{basic_data}). The thin blue line indicates the direct abundance from the 1D spectrum, for reference.}
\label{8005_grad_kpc}
\end{figure*}

\subsection{Radial Metallicity Gradient}

We study the radial gradients of the measured parameters by plotting the parameter value for spaxel (x,y) against the spaxel's distance from the galactic center, in kpc, using the assumed cosmology. In this way, we use the full 2D information on the map analysis for the gradient analysis. We adopt the centroid (white circle in Fig. 9) of the H$\beta$ fluxes as the galaxy center. Using the centroid of the total emission would yield consistent results, as the two centroids differ by less than one spaxel. 

If the two regions with predominantly negative or positive radial velocity observed in Figure~\ref{8005_diagnostic_fbeta_dv_disp_map}, also corresponding to high and low regions of velocity dispersion, belong to the same galaxy, the measured radial metallicity gradient provides a direct constraint on its chemical evolution. 

The radial distribution of the electron temperature (Fig.~\ref{8005_Tegrad}) shows a significant positive correlation with galactocentric distance, indicating the presence of a radial temperature gradient (see Table~\ref{basic_data}). Such a gradient directly implies a radial metallicity gradient, regardless of the abundance tracer employed. Figure~\ref{8005_grad_kpc} presents the radial oxygen gradient of CANDELS~8005, derived from a bootstrap linear fit to data points deriving from S/N $>$3 emission lines used in the diagnostics. The quoted gradient uncertainties in both Figure~\ref{8005_grad_kpc} and Table~\ref{basic_data} correspond to the 17$^{\rm th}$ and 83$^{\rm rd}$ percentiles of the bootstrap distribution.

We find that the oxygen abundance correlates with galactocentric distance, with both Spearman's (r=-0.57, p=1.67$\times$10$^{-5}$) and Pearson's (r=-0.49, p=3.02$\times$10$^{-3}$) tests confirming the statistical significance of the trend. Radial metallicity gradients measured from direct abundances in non-local galaxies are rare, particularly in non-lensed systems (a measurement has been reported for a lensed galaxy by \citet{2026MNRAS.546ag094I}). The gradient of CANDELS~8005 is clearly organized across the full extent of CANDELS~8005. There is one spaxel (41,20) with unrealistic [\ion{O}{3}] flux ratio, indicated with a red circle in Figure~\ref{8005_grad_kpc}, that we exclude it from the gradient measurement, although including the spaxel would not change the science result (see Tab.~\ref{basic_data}, where we give both gradients). Multiplying the errors by 1.28, which is the error underestimation consistent with the discussion on Figure~\ref{8005_O3ratio}, would also not change the gradient slope. The whole analysis, redone without extinction corrections, would also give the same gradient slope within the uncertainties.

Galaxy CANDELS~8005 may indeed be a two-galaxy system, which would be compatible with the shape of the bluer NIRCam images (Fig.~\ref{8005_images}).  The velocity and dispersion maps do not resolve the ambiguity. The mean metallicities of the positive- and negative-velocity spaxels differ only marginally: the average oxygen abundance in the positive-velocity region is higher by $\sim9\%$, corresponding to a statistical significance of $\sim$1$\sigma$ in both O/H (linear) and electron temperature. The density analysis shows a somewhat stronger separation between the two regions, but still below the 3–4$\sigma$ level. With the present data, the distinction between a single system and a pair of merging galaxies remains ambiguous. 

\section{Detailed analysis of CANDELS~7986}
\begin{figure*}[ht]
\centering
\includegraphics[width=\textwidth]{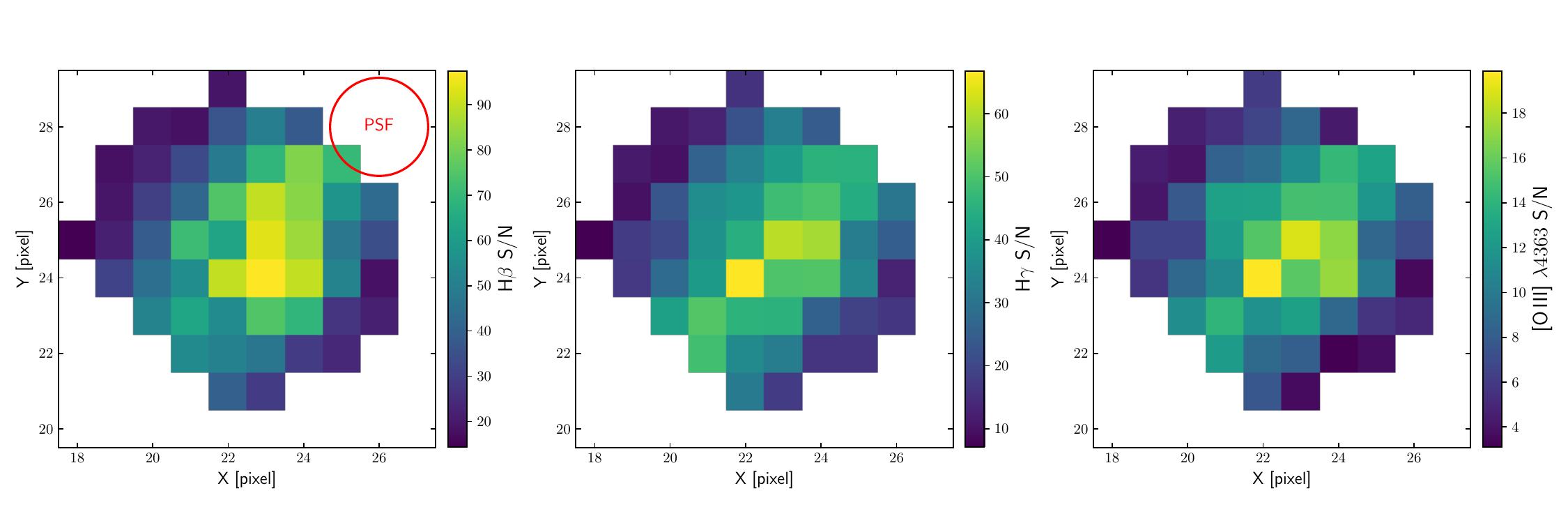}
\caption{CANDELS~7986: As in Fig.\ref{8005_SN_map}, but for CANDELS~7986.}
\label{7986_SN_map}
\end{figure*}

\begin{figure*}[ht]
\centering
\includegraphics[width=\columnwidth]{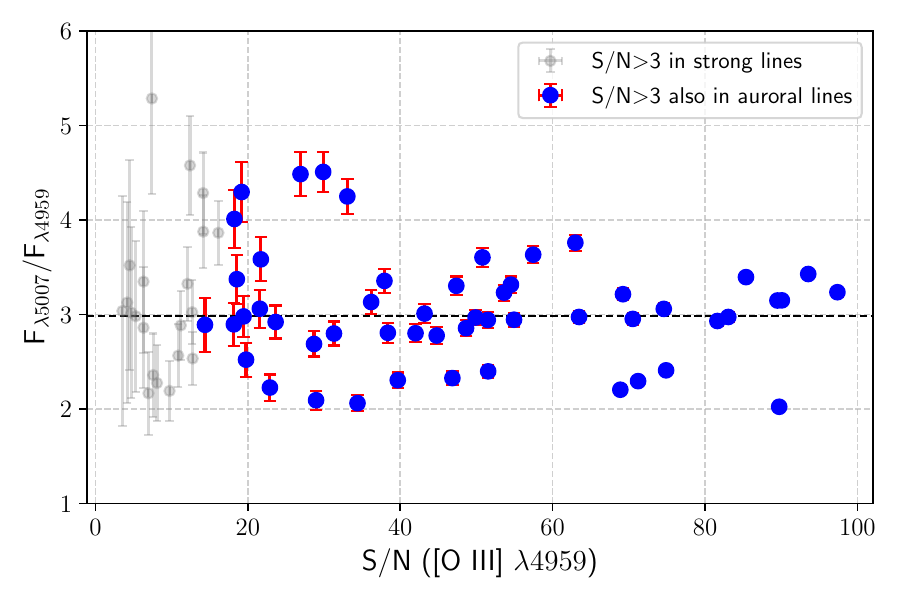}    
\caption{As in Fig.~\ref{8005_O3ratio}, but for CANDELS~7986.}
\label{7986_O3ratio}
\end{figure*}

\begin{figure*}[ht]
\centering
\includegraphics[width=\columnwidth]{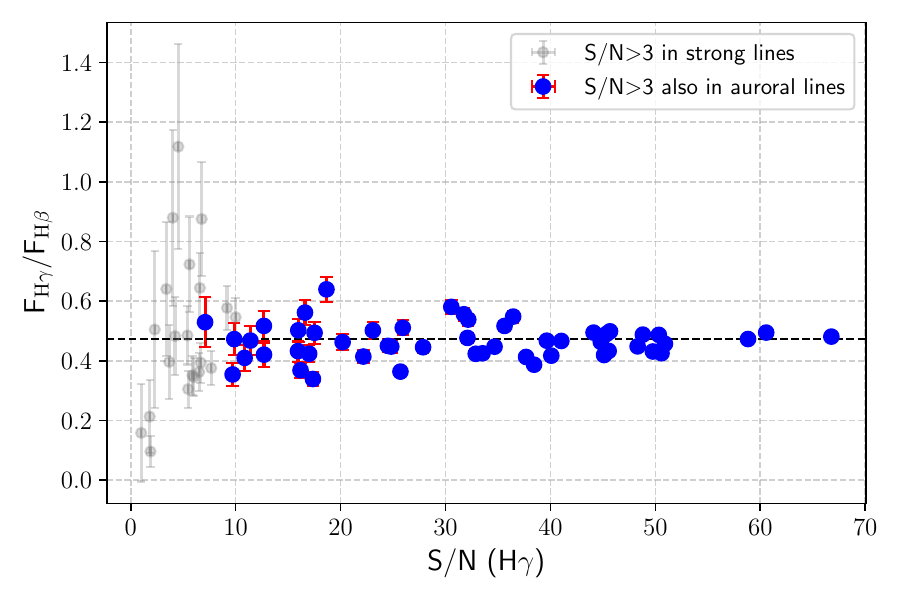}    
\caption{As in fig.~\ref{8005_ext}, but for CANDELS~7986.}
\label{7986_ext}
\end{figure*}
    
\subsection{1D spectral analysis}

The collapsed 1D spectrum of CANDELS~7986 (Figure~\ref{7986_onedspec}) shows the [\ion{O}{2}] doublet, all Balmer series lines from H11 through H$\beta$ (with the exception of H$\epsilon$, which falls in the spectral gap), and the [\ion{Ne}{3}] $\lambda$3869 line (while $\lambda$3968 is also in the gap). The spectrum further exhibits both nebular and auroral [\ion{O}{3}] lines, as well as \ion{He}{1} $\lambda$4471. Weak \ion{He}{2} $\lambda$4686 emission is present, but there is no clear detection of \ion{Ne}{5} $\lambda$3426.

We measure the redshift of the galaxy as a whole by fitting the H$\beta$ and the [\ion{O}{3}]~$\lambda\lambda$4959, 5007 lines, obtaining z=4.702. Direct analysis of the collapsed spectrum yields $T_{\rm e}$=15273$\pm$382 K, $N_{\rm e}$=285$\pm$41 cm$^{-3}$. The nebular continuum is also minimal in this galaxy. Using these plasma parameters we calculate the theoretical H$\gamma$/H$\beta$ ratio under case B (assuming a nebula optically thick to Lyman radiation) from emissivity analysis, is 0.473. The extinction is low, with $c_{\rm \beta}$=0.084$\pm$0.04 (or E(B-V)=0.0489 $\pm$0.0233).

The ionic abundance analysis of the 1D spectrum yields
a total oxygen abundance of $\log(\text{O}/\text{H})+12 = 7.95^{+0.037}_{-0.040}$ (Table~\ref{basic_data}). In this galaxy, the O$^+/$O$^{2+}$ abundance ratio is $\sim10\%$, thus the possible impact of a different zone temperature for O$^+$ would be even lower than in CANDELS~8005.

\begin{figure*}[ht]
\centering
    \includegraphics[width=\textwidth]{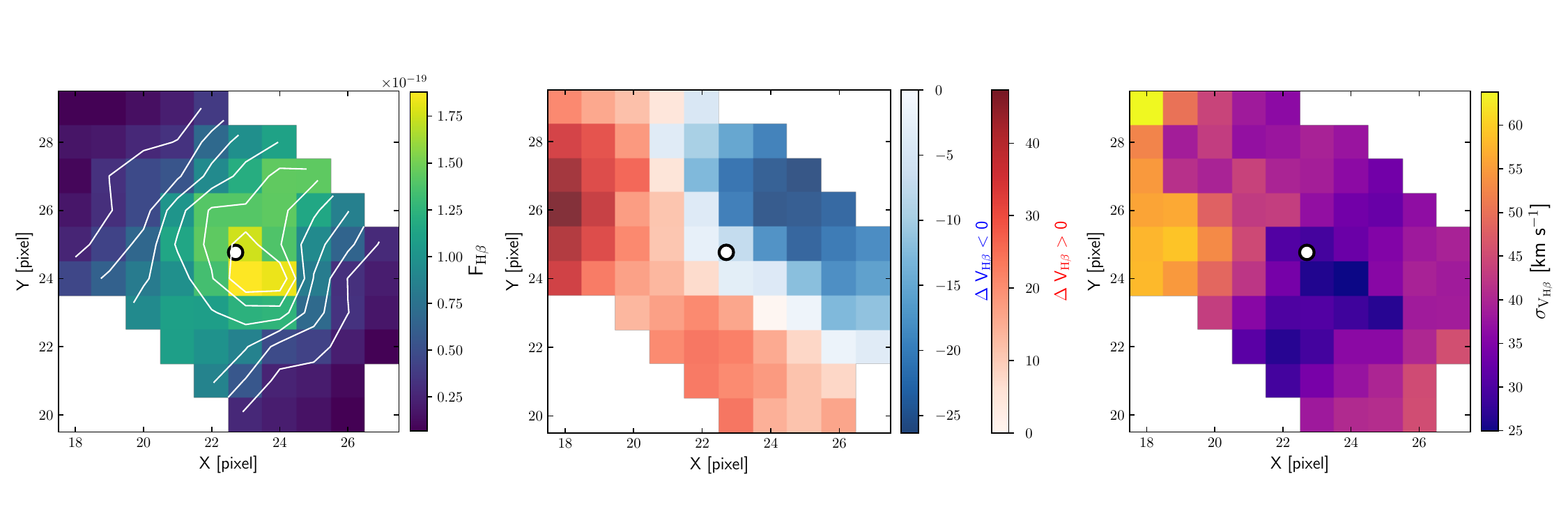} 
 \caption{CANDELS~7986: As in Fig.~\ref{8005_diagnostic_fbeta_dv_disp_map}, but for CANDELS~7986.}
         \label{7986_diagnostic_fbeta_dv_disp_map}
   \end{figure*}

\begin{figure*}[ht]
\centering
\includegraphics[width=\textwidth]{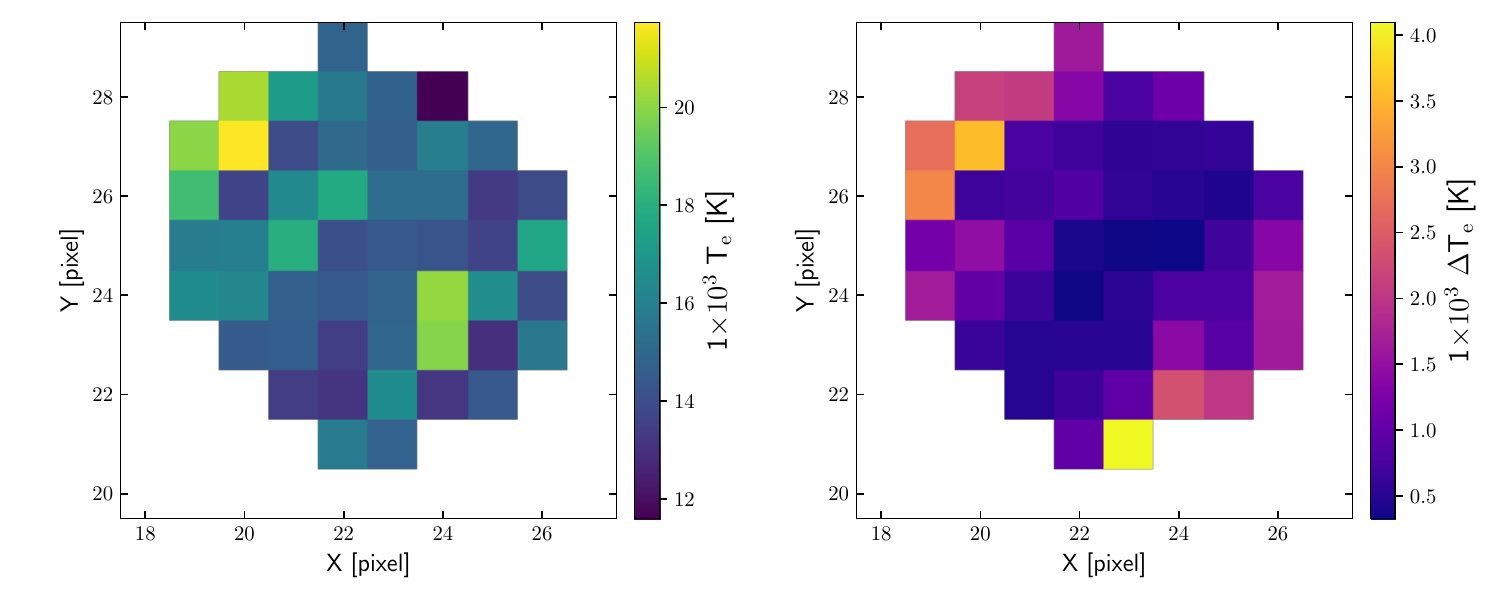}    
\caption{CANDELS~7986: as in Fig.~\ref{8005_diagnostic_Te_map}, but for CANDELS~7986.}
\label{7986_diagnostic_Te_map}
\end{figure*}

\begin{figure*}[ht]
\centering
\includegraphics[width=\textwidth]{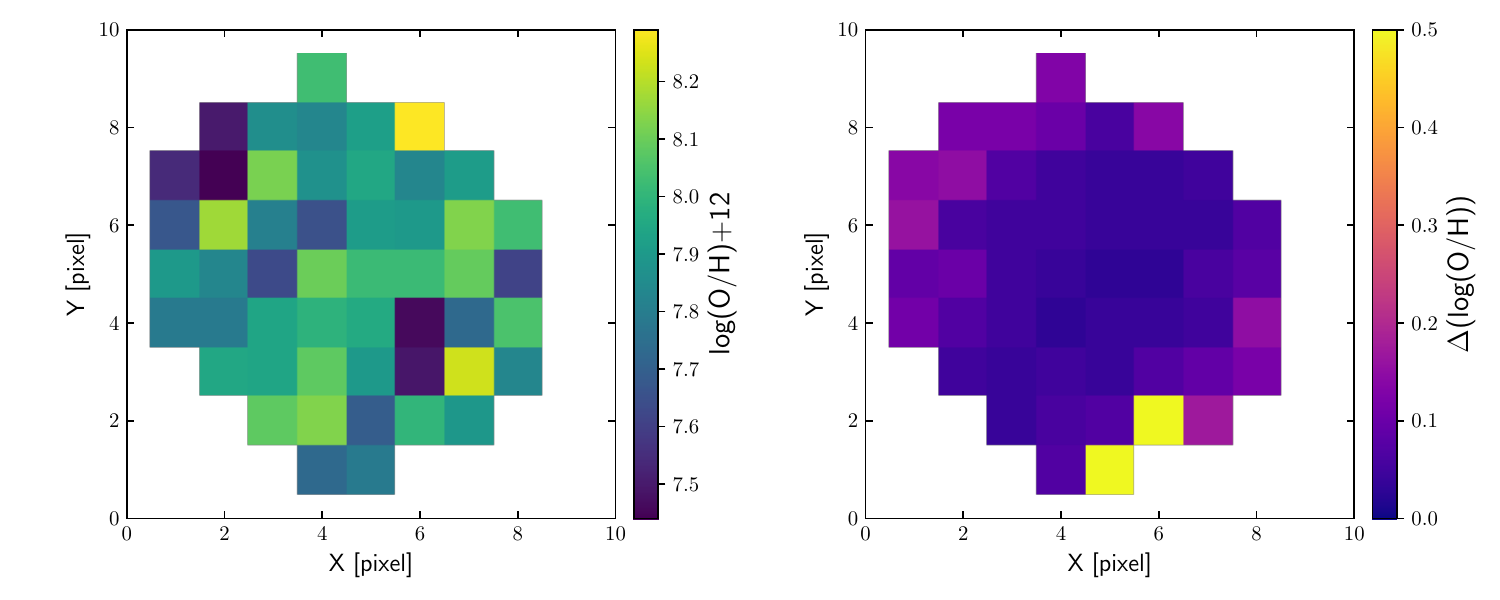}    
\caption{CANDELS~7986: as in Fig.~\ref{8005_diagnostic_OH_map}, but for CANDELS~7986.}
\label{7986_diagnostic_OH_map}
\end{figure*}

\subsection{2D spectral analysis}

The 2D analysis of CANDELS~7986 is based on $10 \times 10$ pixel maps (0.5 $\arcsec$ on a side), similarly to CANDELS~8005, covering the extent of the galaxy seen in the narrow-band image. Restricting the analysis to spaxels with S/N $>$ 3 in all relevant lines including the auroral emission reduces the sample to 51 spaxels. 

Figure~\ref{7986_O3ratio} shows that the [\ion{O}{3}] strong-line ratio deviates more significantly from the theoretical value than in galaxy CANDELS~8005 (here, the reduced $\chi2$ analysis gives 2.95). This may be due to resampling noise \citep[see, e.g.,][their Figure 11]{2023AJ....166...45L} biasing the single-spaxel spectra in our reconstructed data cubes; this effect is expected to be stronger in CANDELS~7986 than CANDELS~8005 since, over the relevant wavelength range, the spectrum is more strongly tilted with respect to the detector rows. Spatial binning would mitigate this effect, but we choose to preserve the spatial resolution of our data.

The extinction by pixel is shown in the Balmer ratio plot (Fig~\ref{7986_ext}). The highest S/N spaxels show minor dust absorption without clear zones of much higher or lower extinction. The lack of correlation between the Balmer ratio and $T_{\rm e}$ further rules out collisional excitation as a significant factor in this galaxy as well as in CANDELS~8005.

 In Figure~\ref{7986_diagnostic_fbeta_dv_disp_map} we show the spatial distribution of the H$\beta$ flux (left panel), the radial velocity measured from the H$\beta$ line fit (central panel), and the velocity dispersion, corrected for instrumental effects (right panel). The H$\beta$ map reveals a compact bright core surrounded by a fainter, diffuse periphery. Velocities, measured relative to the systemic redshift from the integrated 1D spectrum (as for CANDELS~8005), show a gradient along the NE–SW axis, with velocities going from positive in the NE, to negative in the SW parts of the galaxy, as well as substructure at the spaxel level. This could indicate rotation, or could also be an outflow. The gas kinematics of CANDELS~7986 indicate complex dynamics. Similar to CANDELS~8005, the velocity dispersion does not peak in the center. Furthermore, the photometric and kinematic major axes are misaligned, which points to something more complex than  a simple rotating disk. Note that the NIRCam image of the CANDELS~7986 galaxy shows mild elongation primarily in the N-S direction.

Figure~\ref{7986_diagnostic_Te_map} presents the weak-line analysis and the resulting 2D $T_{\rm e}$ distribution, which reveals some spatial structure, indicative of non-uniform metallicity. This is confirmed in Figure~\ref{7986_diagnostic_OH_map}, the metallicity map. Error propagation results in substantial uncertainties in individual spaxels, particularly in the outer regions, which limits the precision of metallicity measurements in the galaxy periphery.

\begin{figure*}[ht]
   \centering
\includegraphics[width=1\columnwidth]{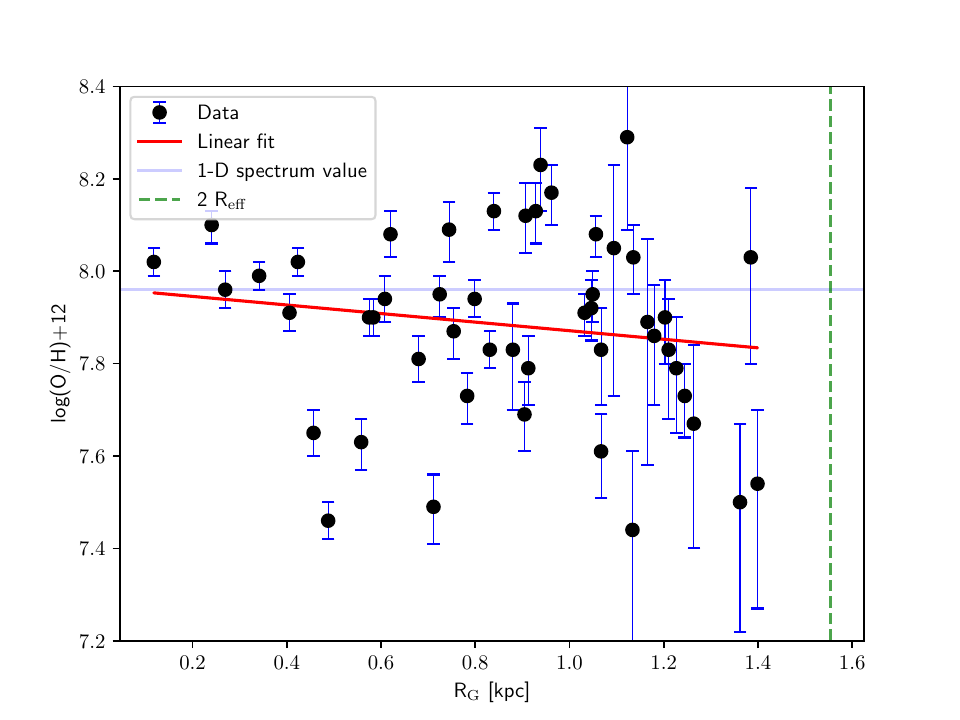}  
      \caption{CANDELS~7986: As in Fig.~\ref{8005_grad_kpc}, but for CANDELS~7986. }
         \label{7986_grad_kpc}
   \end{figure*}

\subsection{Radial Metallicity Gradient}

In CANDELS~7986, the analysis of galactocentric metallicity gradient (Fig.~\ref{7986_grad_kpc}) reveals a slope which is not statistically significant above the $\sim 1\sigma$ level. The scatter in metallicity from spaxel-to-spaxel around the best-fit gradient is larger in this galaxy than in CANDELS~8005. This may be related to the resampling noise effects that also lead to the larger scatter in the \ion{O}{3} doublet ratio in Figure~\ref{7986_O3ratio}.  We further examine whether the oxygen abundances in individual spaxels correlate with the flux intensity (e.g., H$\beta$ surface brightness). While an abundance peak coincides with the brightest region, intermediate-intensity spaxels show a wide range of oxygen abundances. 

\section{Discussion}

\subsection{Comparison between direct and strong-line abundances at high redshift}

\begin{figure*}[ht]
   \centering
\includegraphics[width=1\columnwidth]{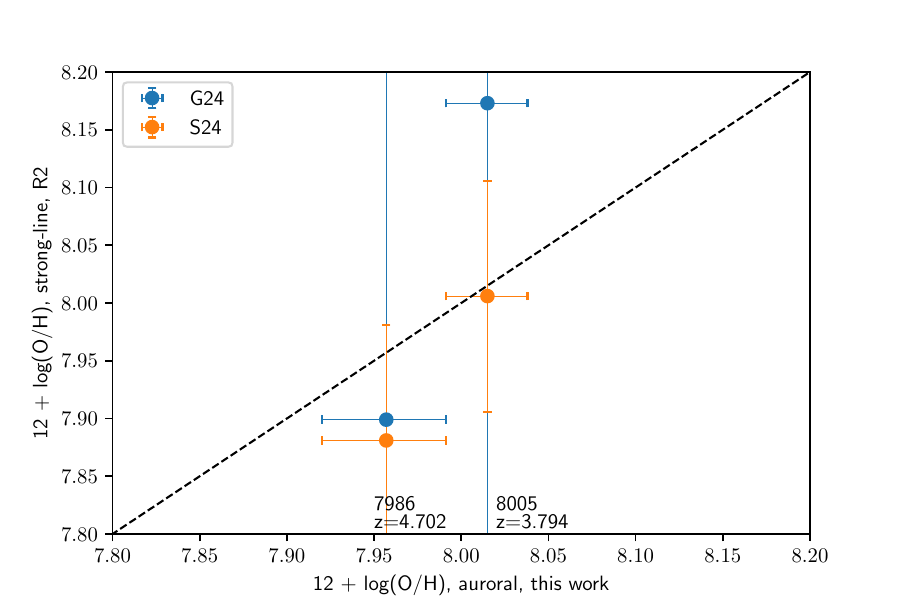}    
      \caption{Comparison between strong-line and direct abundances for the galaxies of this study. Strong-line indices used here are from \citet{2024ApJ...972..113G} (G24), with z=4 for CANDELS~8005 and z=5 for CANDELS~7986, and \citet{2024ApJ...962...24S} (S24). }
         \label{direct_strong_comp}
   \end{figure*}

We calculate 1D strong-line abundances for both galaxies in order to compare them with the direct abundances derived in this work. We adopt the latest strong-line calibrations for high-redshift galaxies from \citet{2024ApJ...972..113G}, which provides various indices for a range of redshifts, and \citet{2024ApJ...962...24S}, where the indices are given for 2$<$z$<$9, and explore the resulting abundances based on the available emission lines, thus on the R2 and R3 indices for both calibrations. The derived values are listed in Table~\ref{basic_data} and plotted against the direct abundances in Figure~\ref{direct_strong_comp}.

We measure the oxygen abundance of CANDELS~8005 from the formulation in \citet{2024ApJ...972..113G}, index R2, z=4, which yields one real root. The R3 index from the same reference does not yield real solutions for this galaxy. Using the indices from \citet{2024ApJ...962...24S}, we also find a single valid solution from the R2 index, and two solutions from the R3 index, one of which yielding the same abundance as the R2 index. We adopt the common root as the representative abundance from \citet{2024ApJ...962...24S}.

For CANDELS~7986, the \citet{2024ApJ...972..113G} R2 index for z=5 also yields one real root, which we adopt. The R3 index from both \citet{2024ApJ...972..113G} and \citet{2024ApJ...962...24S} yields no real solutions. As in 8005, the R2 index formulation from \citet{2024ApJ...962...24S} also yields a valid root.

In Figure~\ref{direct_strong_comp}, we plot the chosen solutions and direct abundances for each galaxy. We observe a good correspondence between direct and strong-line abundances for both galaxies for both index prescriptions, in the case of \citet{2024ApJ...972..113G} when using the model redshift closest to the observed redshift of each galaxy.

While the agreement between direct and strong-line abundance estimates is encouraging, the large uncertainties associated with the strong-line calibrations limit their utility for detailed 2D analysis, although abundances in different spatial areas with relative large uncertainties are still useful for gradient measurements. We compute a strong-line gradient for CANDELS~8005 using the R2-based abundances from \citet{2024ApJ...962...24S}, obtaining $\Delta(\log(\text{O/H}))/\Delta R_{\rm G} = 0.0121\pm$0.1. This value is consistent with the gradient derived from auroral lines within the uncertainties, albeit with significantly lower statistical significance.

\subsection{Star formation versus AGN activity}

 We test whether CANDELS~8005 and CANDELS~7986 are likely to host Active Galactic Nuclei (AGNs), based on the emission lines measured in each galaxy, including the oxygen auroral lines. The Balmer lines for both galaxies are narrow, with no evidence of an AGN broad line component. We use several indices for placement in the modified BPT \citep{1981PASP...93....5B} diagrams that allow galaxy type discrimination. We used two recent schemes of AGN discrimination for JWST spectral data at high redshift \citep{2024A&A...691A.345M,2025arXiv250203519B}, calculating the O3O2 Ne3O2, O33, and O3Hg indices and their uncertainties (see Table~\ref{indices}) from the collapsed spectrum, to be compared with Fig.\ 1 through 3 of \citet{2024A&A...691A.345M} and Fig.\ 5 in \citet{2025arXiv250203519B}. Finally, we also tested the galaxy location on the log($\lambda$3727/H$\beta$) versus log(($\lambda$4959 + $\lambda$ 5007)/H$\beta$) plot, as in \citet[][]{2017sdgh.book.....H} (their Fig. 20). 
 
 We find that both CANDELS~8005 and CANDELS~7986 are unlikely to be hosting AGNs, being in the diagnostic regions occupied by star forming (SF) galaxies. The relatively high values of the intermediate ionization indices (i.e., R3) in both galaxies are consistent with intermediate- to high-ionization SF galaxies, and thus the usage of auroral lines to determine plasma diagnostics and abundances is adequate for the data at hand.

\begin{table*}[ht]
\caption{Index values from the 1D spectra}
\label{fluxes} 
\begin{tabular}{rrll}
\colhead{Index} & \colhead{Expression}& \colhead{CANDELS~8005}& \colhead{CANDELS~7986}\\
\hline

R2& log([\ion{O}{2}]~$\lambda\lambda$3727,29)/H$\beta$)& 
0.076 $\pm$ 0.012&
-0.057 $\pm$ 0.012 \\

R3& log([\ion{O}{3}]~$\lambda$5007/H$\beta$)& 
0.832 $\pm$ 0.002&
0.845 $\pm$ 0.005\\ 

R23& log([\ion{O}{2}]~$\lambda\lambda$3727,29 + [\ion{O}{3}]~$\lambda\lambda$4959, 5007)& 
1.012 $\pm$ 0.012&
1.008 $\pm$ 0.012\\

O3O2& log([\ion{O}{3}]~$\lambda$5007/([\ion{O}{2}]~$\lambda\lambda3727, 29$) & 
0.757 $\pm$ 0.012&
0.902 $\pm$ 0.011 
\\

Ne3O2& log([\ion{Ne}{3}]~$\lambda$3869/([\ion{O}{2}]~$\lambda\lambda3727, 29$)& 
 -0.411 $\pm$ 0.017&
$\dots$
\\

O3H$\gamma$& log([\ion{O}{3}]$\lambda$4363/H$\gamma$)& 
-0.635 $\pm$ 0.022&
-0.525 $\pm$ 0.022\\

O33&   log([\ion{O}{3}]~$\lambda$5007]/[\ion{O}{3}]~$\lambda$4363)& 
1.806 $\pm$ 0.021&
1.710 $\pm$ 0.0212
\\
\hline
\end{tabular}
\label{indices} 
\end{table*}

\subsection{Evolution of the Radial Metallicity Gradients}

The current observational evidence of galactic chemical evolution, based on abundances from individual probes in observed galaxies, supports an inside-out evolutionary scenario, in which the central regions experience earlier star formation and chemical enrichment, leading to older and more metal-rich stellar populations toward galaxy centers \citep[see][]{2013A&A...554A..47G}. Several dynamical processes contribute to mixing and diffusion in the ISM \citep{2019BAAS...51c.161K}. Examples include bar-driven radial mixing \citep{2013A&A...553A.102D}, spiral-arm-induced large-scale streaming motions \citep{2016MNRAS.460L..94G, 2016ApJ...830L..40S}, kiloparsec-scale dilution from the passage of spiral density waves \citep{2015MNRAS.448.2030H}, thermal and gravitational instabilities \citep{2012ApJ...758...48Y, 2015MNRAS.449.2588P}, and interstellar turbulence \citep{2002ApJ...581.1047D,2018MNRAS.475.2236K}. Each of these mechanisms leaves a distinct imprint on the chemical homogeneity of galactic disks.

Given the complexity and strong parameter dependence of these processes, models of galactic chemical evolution require robust observational constraints. Measurements of elemental abundances in spiral galaxies from individual H {\sc ii} regions provide the most direct constraints, particularly through radial metallicity gradients, which can be traced across cosmic time using both high-redshift galaxies and the fossil record preserved in old and young stellar populations in the local Universe.

\begin{figure*}[ht]
\centering
    \includegraphics[width=\textwidth]{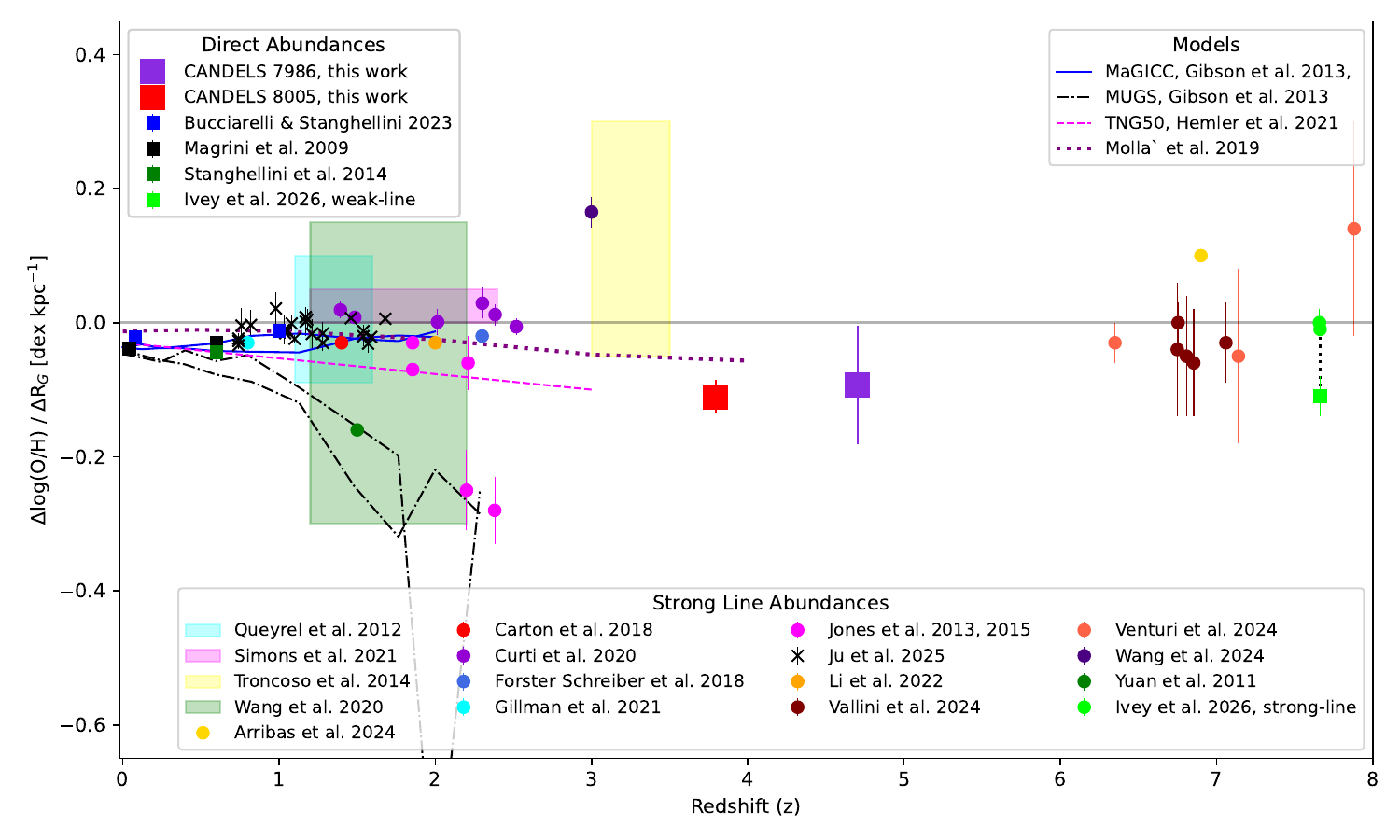}    
      \caption{Radial metallicity gradient slope versus redshift for local and distant galaxies. The data include strong-line abundance gradients (circles and crosses): \citet{2013ApJ...765...48J,2015AJ....149..107J,
2011ApJ...732L..14Y,2012A&A...539A..93Q,2014A&A...563A..58T,2018MNRAS.478.4293C,2018ApJS..238...21F,2020ApJ...900..183W,2020MNRAS.492..821C,2021ApJ...923..203S,2021MNRAS.500.4229G,2022ApJ...929L...8L,2024A&A...691A..19V,2024A&A...688A.146A,2024MNRAS.527...10V,2024AAS...24325101W,2025ApJ...978L..39J}, direct abundance gradients (solid squares): \citet{2023A&A...680A.104B,2009ApJ...696..729M,2014A&A...567A..88S}, and the results by \citet{2026MNRAS.546ag094I}, where the gradients of ID6355 from the strong and weak line methods are connected with a dotted line. We also plot representative chemical evolution models (lines) \citep{2013A&A...554A..47G,2019MNRAS.482.3071M,2021MNRAS.506.3024H}. The direct-abundance gradients from PNe of local galaxies (MW, M33, and M81), measured at z=0, have been placed at the redshift estimated by measuring the look back time of the progenitors of the various PN populations, assuming that there has been no radial migration in the galaxy's lifetime.}
\label{ev_grad}
\end{figure*}

In Figure~\ref{ev_grad} we plot the gradient slopes of CANDELS 8005 and CANDELS 7986, with other direct-abundance gradients available in the literature, versus their redshift. We also mark on the plot a representative sample of strong-line gradients, together with a sample of simulations available in the literature. In order to keep the comparison clear we limit the simulations and the strong-line gradients to a few key examples (see, e.g., \citet{2024A&A...691A..19V} for additional data references). 

Apart from the GARDEN measurements, direct abundances of individual probes have been available to date only for local galaxies. Gradient evolution studies, based on gradient slopes at various redshifts, have been constrained only by strong-line abundances, where in some cases the abundance error bars are of the same magnitude as the gradient slopes.  PNe in local galaxies are also plotted at redshifts whose look-back times correspond to the PN progenitor ages. 
The most massive stars ($<$10 Myr) have a relatively short lifetime, and H~{\sc ii} regions trace the instantaneous chemical abundances in the ISM with their abundances tracing the chemical history of the galaxy since formation.  Instead, PNe are the ionized ejecta of AGB stars, with initial masses in the $\sim$1-8 M$_{\odot}$ range. Since AGB stars do not process, or minimally process, oxygen or other $\alpha$-elements, their oxygen abundances are unchanged since the time of the progenitor's formation. 
Depending on their stellar progenitor mass, PN abundances can help to explore look-back times from $\sim$1--8 Gyr. It is understood that radial migration may affect the gradients of old stellar populations, a factor to add to the uncertainties of the PN data points at z$>$0.

As seen in Figure~\ref{ev_grad}, radial oxygen gradients of H~{\sc ii} regions and PNe with high mass progenitors in local galaxies are shallow and negative, while gradients for the same galaxies but from older PN populations -- i.e., corresponding to z$\sim$0.5-1 -- tend to be closer to 0.  All gradients at moderate redshifts (z$<$1) seem to indicate that radial metallicity gradients steepen with the time evolution of galaxies. However, this scenario consists of significant oversimplifications and hinges on several assumptions. Key among these are the notions that radial migration of older stellar populations is minimal, as seen in nearby galaxies, or that any such migration is smaller than the uncertainties associated with distance measurements.

Around Cosmic Noon ($1 \lesssim z \lesssim 3$), strong-line gradients span a wide range, while at very high redshifts ($z>5$) gradients from strong-line analyses tend to be negative and shallow. The only direct-method gradient measurements for $z>1$, excluding lensed galaxies, come from this study: a well-measured negative gradient for CANDELS~8005 ($z=3.794$) and an uncertain slope for CANDELS~7986 ($z=4.702$). Considering solely the direct abundances, we conclude that the radial oxygen gradients of star-forming galaxies flatten with time from high redshift through Cosmic Noon, then mildly steepen from Cosmic Noon to $z=0$. Despite being based on a few and heterogeneous data points, this evolutionary trend is broadly consistent with inside-out models of galaxy chemical evolution, such as those presented by \citet{2021MNRAS.506.3024H}.

The broad discussion of metallicity gradient evolution at this time, when the direct abundance gradients measured are so few, should include the strong-line measurements. Strong-line gradients indicate a complex scenario at Cosmic Noon, where we see a large range of gradient slopes; they also show that at very high redshifts gradient slopes seem to be again confined to be negative and shallow, as in the nearby Universe. Furthermore, comparing galaxy gradients tends to oversimplify the fact that the observed galaxies may be different in nature. In particular, it is observed that gradient slopes of local galaxies become more negative with stellar mass in the log(M/M$_{\odot})$ range of 8.5-10.4 \citep{2025A&A...693A.150K}

Given these complexities, we refrain from drawing definitive conclusions about the chemical evolution of galaxies in this paper. With this pilot experiment we assert that spatially-resolved direct abundances are increasingly attainable with current technology, potentially unlocking groundbreaking insights and posing significant challenges to our understanding of the chemical evolution of galaxies.

\section{Conclusions, and Future Endeavors}

In this study, we successfully measured the direct, spatially resolved metallicity distribution of two high-redshift galaxies. A statistically significant radial oxygen abundance gradient of $-$0.111$^{+0.026}_{-0.025}$ [dex/kpc] was determined for CANDELS~8005, while the data are consistent with no gradient found in CANDELS-7986. Our analysis revealed distinct kinematic features, with evidence suggesting that CANDELS~8005 may comprise two merging systems. We find no indication of active galactic nucleus (AGN) activity in either CANDELS~7986 or CANDELS~8005.

We compared the direct abundances with strong-line estimates, finding good agreement with model calibrations at the relevant redshifts. This consistency reinforces the reliability of our methods and adds confidence to the derived abundances.

Our results validate the use of the GARDEN slitlet-stepping technique to obtain two-dimensional spectra of high-redshift galaxies, enabling detailed inspection of their chemical and ionization properties. Future work will expand upon this by exploring the kinematics and dynamics of these systems over larger spatial extents, with a particular focus on the internal motions within CANDELS~8005.

In addition, we will conduct a comprehensive strong-line abundance analysis across the full GARDEN sample, applying consistent diagnostics wherever the required emission lines are available. This will facilitate a broader statistical and physical characterization of the chemical enrichment and evolutionary states of high-redshift galaxies.

\begin{acknowledgments}

We are grateful to the referee for the thorough and thoughtful review, and for the constructive comments that helped strengthen the manuscript. We warmly thank Christophe Morisset for his suggestions on the PyNeb routine usage. This work is based on observations made with the NASA/ESA/CSA \textit{James Webb Space Telescope}, obtained at the Space Telescope Science Institute, which is operated by the Association of Universities for Research in Astronomy, Incorporated, under NASA contract NAS5-03127. Support for program number GO-2123 was provided through a grant from the STScI under NASA contract NAS5-03127. The data were obtained from the Mikulski Archive for Space Telescopes (MAST) at the Space Telescope Science Institute.   These observations are associated with program \#2123, and can be accessed via MAST. This work also uses data from the JWST Advanced Extragalactic Survey (JADES) data release DR2, consisting of data from GTO programs 1180, 1181, 1210 and 1286, and GO programs 1895, 1963, and 3215.  
JADES DR2 can be accessed via doi: 10.17909/z2gw-mk31 \citep{jades_2023_data}.

\end{acknowledgments}

\vspace{5mm}
\facilities{JWST(NIRSpec)}

\software{PyNeb
          }

\bibliography{biblio.bib}{}
\bibliographystyle{aasjournal}

\end{document}